\documentclass[prd,aps,floatfix,superscriptaddress,nofootinbib,twocolumn,10pt,English]{revtex4}

\usepackage{amssymb,amsmath}
\usepackage{graphicx}
\usepackage[percent]{overpic}
\usepackage{overpic}
\usepackage{color}
\usepackage[colorlinks=true]{hyperref}
\numberwithin{equation}{section}

\begin{document}  

\title{Inspiralling eccentric binary neutron stars: orbital motion and tidal resonance}

\author{ Huan Yang}
\affiliation{University of Guelph, Guelph, Ontario N2L3G1, Canada}
\affiliation{Perimeter Institute for Theoretical Physics, Waterloo, Ontario N2L2Y5, Canada}

\begin{abstract} 
We study the orbital evolution of eccentric binary neutron stars. The motion  follows a Quasi-Keplarian orbit with perturbations due to tidal couplings. We find that the tidal interaction between stars contributes to orbital precession in addition to the Post-Newtonian procession. The coupling between the angular and radial motion of the binary also excites a series of harmonics in the stars' oscillation. In the small eccentricity limit, this coupling mainly gives rise to an additional orbital resonance, with the orbital frequency being one third of the f-mode frequency.
For a binary with initial eccentricity $\sim 0.2$ at $50$Hz orbital frequency, the presence of this tidal resonance introduces $\sim \mathcal{O}(0.5)$ phase shift in the gravitational waveform till merger, subject to uncertainties in neutron star equation of state and the distribution of binary component masses. 
Such phase shift in the late-inspiral stage is likely detectable with third-generation gravitational-wave detectors.
\end{abstract}

\maketitle

\section{Introduction}

Neutron star  mergers produce copious gravitational wave (GW) and electromagnetic (EM) radiation, that encode critical information about state of matter under the extreme physical conditions 
(gravity, pressure, temperature, etc.).  
In particular, the detection of GW170817 \cite{TheLIGOScientific:2017qsa} and subsequent observations from gamma-ray band to radio-wave frequencies has ushered in a new era of multi-
messenger astronomy \cite{abbott2017multi,abbott2017gravitational,abbott2017estimating,coulter2017swope,troja2017x,2041-8205-848-2-L21,kochanek1993gravitational,li1998transient}. As the sensitivity of current-generation GW detectors degrades significantly above $1$kHz \cite{Miao:2017qot,martynov2019exploring}, directly probing the post-merger GW signal from binary neutron 
stars (BNSs) is less likely in a few years. The inspiral part of the GW signal, however, not only provides information about the binary orbital parameters, such as component masses, eccentricity, 
distance, etc., but also the matter properties through the measurement of tidal deformability and possible resolution of the merger frequency. For example, the tidal love number constraint from GW170817 
has greatly refined the possible parameter range of NS EOS.

Most of the compact binary mergers are believed to be (nearly) circular \footnote{A recent publication has argued that Post-Newtonian radiation reaction excites small eccentricities $\mathcal{O}
(0.01)$ in the late inspiral stage \cite{Loutrel:2018ydu}, which appears to be interesting to be the discussion in this work.}. However, recent studies have shown the possibility of forming eccentric 
compact binaries in the LIGO (Laser Interferometric Gravitational Wave Observatory) band by resonant and hierarchical triple and quadruple systems in globular clusters \cite{2003ApJ...598..419W,2013PhRvL.111f1106S,naoz2016eccentric,2017ApJ...841...77A,2018arXiv180508212R,samsing2017assembly,samsing2018eccentric,liu2018enhanced,hoang2018black}, in addition to dynamically 
captured binaries \cite{east2013observing}. These events may be rare comparing to circular inspirals, but they contain important information about their formation channel(s), environment and distribution, through the spin and 
eccentricity measurement. For eccentric binary neutron stars, we shall show that the coupling between radial and angular orbital motions give rises to new tidal resonances, that can be possibly 
detected by  third-generation GW detectors such as Einstein Telescope \cite{punturo2010einstein} and Cosmic Explorer \cite{abbott2017exploring}.

The motion of inspiring eccentric binary black holes (BBHs), within the Post-Newtonian framework, can be described by the Quasi-Keplerian (QK) orbits  \cite{blanchet2014gravitational}. At the zeroth order, the QK orbit coincides 
with the Newtonian elliptical orbit. With higher order Post-Newtonian effect included, the radial motion generically oscillates with a different period from the angular motion, which is known as the general relativistic precession. In addition, radial and angular motions also receive 
periodic corrections that can be expanded in a Post-Newtonian manner. Alternatively, it is possible to utilize the Effective-One-Body (EOB) framework, which resembles the Post-Newtonian expansion in the EOB spacetime \cite{Hinderer:2017jcs}. Based on these frameworks, different waveform models have been developed to characterize the GW signature of BBHs, which have achieved 
in decent accuracy for low-medium eccentricity  binaries \cite{huerta2018eccentric,huerta2017complete,cao2017waveform,hinder2018eccentric}, with promising potential to realize fast frequency-domain waveform model with arbitrary eccentricities \cite{Huerta:2016rwp,moore2018fourier}. 


For eccentric  BNSs, the tidal coupling between stars introduces extra complication in the equation of motion. For example, in \cite{Yang:2018bzx} we have studied highly eccentric BNS in the inspiral 
stage, where the f-modes of NSs are excited during the close pericenter passages, and subsequently play important roles in the orbital evolution. With sufficiently accurate orbit model, it might be possible to combine f-mode emission from different pericenter encounter cycles to boost the signal-to-noise ratio of detection, e.g., with the coherent stacking method \cite{yang2017black,PhysRevD.97.024049,berti2018extreme}. While the formation of such systems may 
require rather restrictive initial parameters, it is still important to characterize the tidal effect in medium-low eccentricity BNSs, in addition to the understanding of equilibrium and dynamic tide in the circular 
orbit limit \cite{Steinhoff:2016rfi,Flanagan:2007ix}.

In this work, we adopt the QK orbit as the unperturbed solution without the tidal effect, and compute the NS oscillation in response to the orbital motion and tidal couplings.
For eccentric orbits, the tidal budge induced on a star generally does not point to the companion star in the binary. As a result, the binary orbital angular momentum continuously exchanges with the NS mode angular momentum 
within orbital timescales. These NS oscillations also back-react on the binary orbit, giving rise to perturbations of the QK orbit that affect the GW radiation. In the low eccentricity limit, we explicitly 
evaluate these perturbation terms and determine their influence on the orbital phases in the time domain. Although the discussion of the NS mode and the orbital dynamics has been specified to 
BNSs, the result is also applicable to BH-NS binary \cite{Yang:2017gfb}, with the oscillation of the BH set to be zero. The construction of a frequency-domain waveform model will be left to further work.

This paper is organized as follows. In Sec.~\ref{sec2} we present the basic formalism to describe the motion of an eccentric BNS system under the influence of tidal couplings. In Sec.~\ref{sec:new} we apply this formalism to the Newtonian orbits, characterize the importance of tidal effects
for low-eccentricity systems and discuss the tidal resonance effect. In Sec.~\ref{sec4} we repeat the exercise for Post-Newtonian orbits. We conclude in Sec.~\ref{sec5}. Natural units with $c=1, G=1$ are used through out the analysis.

\section{Basic Formalism}\label{sec2}

The Hamiltonian of a BNS system, including the leading-order tidal excitations of the stars, can be expressed as 
\cite{Steinhoff:2016rfi,Flanagan:2007ix}
\begin{align}\label{eq:h}
\mathcal{H}& = \mathcal{H}_{\rm orb} + \sum_{n} \frac{(\dot{Q}^n)^2+\omega^2_n (Q^n)^2}{2} +\epsilon \frac{\mathcal{E}^{ij}Q^n_{ij}}{2}\,\nonumber \\
& = \mathcal{H}_{\rm orb}+\mathcal{H}_{\rm mode}+\mathcal{H}_{\rm int}\,,
\end{align}
where the tidal field is labeled with $\epsilon$    for book keeping purpose, and $\omega_n$ is the eigenfrequency for mode $n$.
In stellar perturbation theory \cite{unno1979nonradial} the modes of a three-dimensional star are often indexed by the angular nodal number $\ell$, azimuthal nodal number $m$ and radial  nodal number $n_r$. Here $n$ represents the collection of these three indices: $n=\{\ell m n_r\}$.
In particular, the gravitational response of f-mode generally dominates over other modes (e.g., p-modes and g-modes) of the NS, so that in this work we primarily focus on
the f-mode excitations. The electric part of the tidal tensor, $\mathcal{E}_{ij}$,  can be evaluated based on the relative displacement of the binary \cite{poisson2010geometry,Yang:2018bzx}, and the local spacetime of the target star is influenced by the tidal environment generated by its companion (also see the application in extreme mass-ratio inspiral systems in \cite{yang2017general,bonga2019tidal}).
The orbital Hamiltonian $ \mathcal{H}_{\rm orb}$, expanded in the Post-Newtonian format, can be found in \cite{Memmesheimer:2004cv} up to the $3$rd Post-Newtonian order.
Because of the centre-of-mass conservation, the orbital motion is fully characterized by the relative distance and orbital angle,
which orginally motivated the development of the EOB formalism \cite{damour2014general}. The modal quadrupole moment $Q^n_{ij}$  induced by the tidal field,  and the modal
excitation displacement $Q^n$,
are related to each other through
\begin{align}\label{eq:qdecom}
Q_{ij} =\sum_n \mathcal{O}^n_{ij} Q_n\,,
\end{align}
with the overlap tensor defined as
\begin{align}\label{sec:onij}
\mathcal{O}^n_{ij} =\int \rho d^3 x (\xi^n_i x_j +\xi^n_j x_i)
\end{align}
where $\rho$ is the local density within the star and ${\bf \xi}^n$ is the displacement vector associated with the eigenmode wavefunction, that can be expressed in vector spherical harmonics:
\begin{align}\label{eqeigenwf} 
{\bf \xi}^{(n)} = \left ( \xi^{(n)}_R(r) {\bf
e_r}+\xi^{(n)}_S(r) r \nabla \right ) Y_{lm}(\theta,\phi)\,.
\end{align} 
The determination of $\xi^{(n)}_R$ and $\xi^{(n)}_S$ is
discussed in Appendix A of \cite{Yang:2018bzx}.
Notice that Eq.~\eqref{eq:qdecom} can be derived by combining Eq.~\eqref{sec:onij} the expansion
\begin{align}
{\bf \xi} = \sum_n Q_n {\bf \xi}^n\,,
\end{align}
and it can be further represented as
\begin{align}
\mathcal{O}^n_{ab} = Q^n_\xi \int d \Omega ({\bf e}_a \cdot {\bf e}_r) ({\bf e}_b \cdot {\bf e}_r)\,.
\end{align}
with
\begin{align}
Q^n_\xi = 2 \int^{R_*}_0 dr r^{\ell+1} \rho [\xi^{(n)}_R +(\ell+1) \xi^{(n)}_S]\,.
\end{align}
At this point, all variables are still real. But later on when we talk about specific modes, it is often convenient to use complex wave functions for decomposition. In those cases, $Q_n$ can be complex-valued, although the total displacement $\xi$ remains real.

\subsection{Orbital description}

Under the influence of tidal field, the binary orbit is no longer eccentric, even in the Newtonian limit. However, as the tidal interaction is generally weaker than the direct point-mass gravitational 
attraction, we can expand the orbit as power laws in the tidal-coupling coefficient $\epsilon$
\begin{align}\label{eq:expand}
{\bf x} &={\bf x}_{\rm orb} +\epsilon^2 \delta {\bf x}+\mathcal{O}(\epsilon^3)\,,\nonumber \\
{\bf p} &={\bf p}_{\rm orb} +\epsilon^2 \delta {\bf p}+\mathcal{O}(\epsilon^3)\,,
\end{align}
where ${\bf x}, {\bf p}$ follow an eccentric orbit in the Newtonian description or QK orbit in the Post-Newtonian description, and the orbital evolution is determined by the conserved energy  and angular momentum $E, J$. 
\begin{align}
E & = \mathcal{H}_{\rm orb}({\bf x}_{\rm orb},{\bf p}_{\rm orb})\,, \nonumber \\
J & = \mathcal{J}_{\rm orb}({\bf x}_{\rm orb},{\bf p}_{\rm orb})\,,
\end{align}
where $\mathcal{J}_{\rm orb}$ can also be found in \cite{Memmesheimer:2004cv} up to the 3rd Post-Newtonian order. The perturbation starts at the second order in $\epsilon$ as $Q$ starts at linear order in $\epsilon$, so that the back-reaction of  $Q$ starts at the second order. As the binary orbit decays due to GW radiation  generated by orbital motion and mode oscillation, we have
\begin{align}\label{eq:rd}
 P^E_{\rm orb} + P^E_{\rm mode} +P^E_{\rm \times}&=\dot{E}_{\rm orb} + \dot{E}_{\rm mode} + \dot{E}_{\rm int}\,\nonumber \\
 & = \dot{E} +\left [\epsilon^2 \frac{\partial \mathcal{H}_{\rm orb}}{\partial {\bf x}} \delta x \right ]_{,t}+ \epsilon^2 \left [\frac{\partial \mathcal{H}_{\rm orb}}{\partial {\bf p}} \delta p \right ]_{,t} \nonumber \\
 &+\epsilon^2\dot{E}_{\rm mode} + \epsilon^2\dot{E}_{\rm int} \nonumber \\
 &= \dot{E} +\frac{\partial \mathcal{H}_{\epsilon^2}}{\partial E}\dot{E}+\frac{\partial \mathcal{H}_{\epsilon^2}}{\partial J} \dot{J} \,,\nonumber \\
 P^J_{\rm orb} + P^J_{\rm mode} +P^J_{\rm \times}&=\dot{J}_{\rm orb} + \dot{J}_{\rm mode} \nonumber \\
 & = \dot{J} +\left [\epsilon^2 \frac{\partial \mathcal{J}_{\rm orb}}{\partial {\bf x}} \delta x \right ]_{,t}+ \epsilon^2 \left [\frac{\partial \mathcal{J}_{\rm orb}}{\partial {\bf p}} \delta p \right ]_{,t} \nonumber \\
 &+\epsilon^2\dot{J}_{\rm mode}   \nonumber \\
 &=\dot{J} +\frac{\partial \mathcal{J}_{\epsilon^2}}{\partial E}\dot{E}+\frac{\partial \mathcal{J}_{\epsilon^2}}{\partial J} \dot{J}\,,
\end{align}
where different components of the energy and angular momentum flux $P^{E,J}$ are discussed in Sec.~\ref{sec:ra}. To evolve the orbit, we need to relate them to the secular change of $E$ and $J$. This means that we need to solve the equation of motion for $\delta {\bf x}, \delta {\bf p}, Q$ as a function of $E, J$ and time $t$. Based on the Hamiltonian, the equations of motion are
\begin{align}\label{eq:eom}
&\ddot{Q}^n +\gamma \dot{Q}^n+\omega^2_n Q^n = - \epsilon \,  \mathcal{E}_Q({\bf x}_{\rm orb},{\bf p}_{\rm orb}) \,,\nonumber \\
& \dot{\bf \delta p} = -\frac{\partial^2 \mathcal{H}_{\rm orb}}{\partial {\bf x}\partial {\bf x}} {\bf \delta x} -\frac{\partial^2 \mathcal{H}_{\rm orb}}{\partial {\bf x}\partial {\bf p}} {\bf \delta p} -\frac{\partial \mathcal{H}_{\rm int}}{\partial {\bf x}} \,,\nonumber \\
& \dot{\bf \delta x} = \frac{\partial^2 \mathcal{H}_{\rm orb}}{\partial {\bf p}\partial {\bf x}} {\bf \delta x} +\frac{\partial^2 \mathcal{H}_{\rm orb}}{\partial {\bf p}\partial {\bf p}} {\bf \delta p} \,.
\end{align}
where $\gamma$ is an infinitesimal positive damping rate due to dissipations in the star that have not been accounted for in the Hamiltonian formalism, and $\mathcal{E}_Q := \mathcal{E}_{ij} \mathcal{O}^*_{ij}/2$. Notice that we are taking complex conjugation for $\mathcal{O}_{ij}$ here because the complex wave function is being used. The solution of $Q$ is
\begin{align}\label{eq:dtide}
Q^n = &  -\epsilon / \omega_n \int^t e^{-\gamma(t-t')} \sin \omega_n (t-t') \mathcal{E}_{Q}({\bf x}_{\rm orb},{\bf p}_{\rm orb},t') dt'\, \nonumber \\
&+ Q^n_{\rm init} e^{- i \omega_n t-\gamma t} = Q^n_{\rm driven}+Q^n_{\rm free}\,,
\end{align}
where $Q^n_{\rm init}$ corresponds to the initial value of $Q^n$.

Like a normal weakly-damped harmonic oscillator, $Q^n$ generically contains a ``driven" part that is proportional to $\epsilon$ and a ``free" that satisfies the homogeneous equation of motion 
(the left hand side of the first line of Eq.~\eqref{eq:eom}). Under a periodic driving force with slowly varying frequency $\omega/(2\pi)$ ($\dot{\omega}/\omega \ll \omega$), the driven part $Q_{\rm driven}$ can often be approximated by its adiabatic value
\begin{align}\label{eq:qad}
Q^n_{\rm driven} \approx Q^n_{\rm ad} = -\epsilon \frac{\mathcal{E}_Q (t)}{\omega^2_n -2 i \gamma \omega -\omega^2}\,.
\end{align}

On the other hand, if the initial oscillation is zero or the initial time of integration is far in the past, such ``free" part may be neglected.
However, as we shall see later, when the system evolves across  mode resonances, certain level of free mode oscillation will be excited as well.

\vspace{0.5cm}

\subsection{The radiative terms}\label{sec:ra}

The radiation terms that appear in Eq.~\eqref{eq:rd} can be evaluated from the quadruple formula:
\begin{align}
P^E & = -\frac{1}{5} \langle \dddot{\mathcal{I}}_{jk} \dddot{\mathcal{I}}_{jk} \rangle\,,\nonumber \\
P^J_i & = -\frac{2}{5} \epsilon_{ijk} \langle \ddot{\mathcal{I}}_{jm} \dddot{\mathcal{I}}_{km} \rangle\,.
\end{align}
Because we have assumed both NSs to be non-spinning, their orbital motion should still remain on a plane with the presence of Post-Newtonian correction and tidal effects. 
As a result, the orbital angular momentum is orthogonal to the orbital plane, so that only the normal component (defined as the $z$ direction) of $P^J$ is non-vanishing, and hereafter we drop 
the vector indices of $P^J$.

The orbital energy and angular momentum flux $P^E_{\rm orb}, P^J_{\rm orb}$ are derived in \cite{arun2008inspiralling} up to the third Post-Newtonian order.
The modal fluxes $ P^E_{\rm mode}, P^J_{\rm mode}$  are given by

\begin{align}
P^E_{\rm mode} & =- \frac{1}{5}  \langle |\sum_m \dddot{Q}_m \mathcal{O}^m_{ij}  |^2\rangle \nonumber \\
& = -\frac{8 \pi}{75} Q^2_\xi \langle \sum_m |\dddot{Q}_m  |^2\rangle\nonumber \\
& = -\frac{4 \pi M^2_* Q^4_\xi n^6}{75 q^6} \sum_m W^2_{2m} \sum_k \frac{k^6( |c^m_{3,k}|^2+|s^m_{3,k}|^2)}{[(kn)^2-\omega^2_f]^2}\,,
\end{align}
and \cite{Yang:2018bzx}
\begin{align}
P^J_{\rm mode} & = -\frac{2}{5} \epsilon_{ijk} \langle \sum_m \ddot{Q}_m \mathcal{O}^m_{j h} \sum_{m'} \dddot{Q}_{m'} \mathcal{O}^{m'}_{k h}  \rangle \nonumber\\
& =  -\frac{i 16 \pi}{75} Q^2_\xi \langle \ddot{Q}_2 \dddot{Q}_{-2} -  \ddot{Q}_{-2} \dddot{Q}_{2}\rangle \nonumber \\
& = -\frac{32 \pi}{75} \frac{ W^2_{22} M^2_* }{ q^6}Q^4_\xi n^5 \sum_k \frac{k^5 s^{2}_{3,k} c^{-2}_{3,k}}{[(kn)^2-\omega^2_f]^2}\,,
\end{align}
where we have used Eq.~\eqref{eq:qm} and the fact that only the $\ell=2, m=0,\pm 2$ f modes are relevant for the discussions here (The $m=\pm 1$ modes are not excited \cite{press1977formation,Yang:2018bzx}). These expressions can be made more explicit with the prescription of an unperturbed orbit. 

The cross terms $P^E_{\rm \times}$ and $P^J_{\rm \times}$ comes from the beating between radiation from the orbital motion and the mode excitation. 

\begin{widetext}
\begin{align}
P^E_{\rm \times}  = & \,-\frac{2}{5}  \langle \sum_m \dddot{Q}_m \mathcal{O}^m_{ij} \dddot{I}_{ij}  \rangle \,, \nonumber\\
& =\frac{4 \pi}{25}  \mu \frac{M^{3/2}}{q^{5/2}}\frac{  M_* Q^2_\xi}{ q^3} n^3 \sum_k k^3\frac{(4 s^2_{3,k} -0.5 e s^3_{2,k}+0.5 e s^1_{2,k} )c^2_{3,k} +(4 c^2_{3,k} -0.5 e c^3_{2,k} +0.5 e c^1_{2,k}) s^2_{3,k}}{(k n)^2-\omega^2_f} \nonumber \\
&-\frac{8 \pi}{75}  \mu \frac{M^{3/2}}{q^{5/2}} e  \frac{  M_* Q^2_\xi}{ q^3} n^3 \sum_k k^3 \frac{s^1_{2,k} c^2_{3,k}}{(k n)^2-\omega^2_f}\,,
\end{align}
and
\begin{align}
P^J_{\rm \times}  = &  -\frac{2}{5} \epsilon_{ijk} \langle \ddot{I}_{j h} \sum_{m'} \dddot{Q}_{m'} \mathcal{O}^{m'}_{k h} \rangle 
 -\frac{2}{5} \epsilon_{ijk} \langle \sum_m \ddot{Q}_m \mathcal{O}^m_{j h} \dddot{I}_{k h} \rangle \nonumber \\
  = & \frac{4 \pi}{25} \frac{ \mu M}{q} \frac{  M_* Q^2_\xi}{ q^3} n^3 \sum_k k^3 \frac{[c^2_{1,k}+c^2_{2,k} -e (c^3_{1,k}-c^1_{1,k})+e^2/4(c^4_{0,k}- 2 c^2_{0,k})] s^2_{3,k} +(c \leftrightarrow s)}{(k n)^2-\omega^2_f} \nonumber \\
 & +\frac{4 \pi}{25} \frac{ \mu M^{3/2}}{q^{5/2}}  \frac{  M_* Q^2_\xi}{ q^3} n^2 \sum_k k^2 \frac{(4 c^2_{3,k}-0.5 e c^3_{2,k} +0.5 e c^1_{2,k})c^2_{3,k}+(4 s^2_{3,k} -0.5 e s^3_{2,k}+0.5 e s^1_{2,k})s^2_{3,k}}{(k n)^2-\omega^2_f}\,.
 \end{align}
 \end{widetext}
In the Newtonian limit, $\ddot{I}_{ij}$ and $\dddot{I}_{ij}$ can be found in Eq.(12.77) of \cite{poisson2014gravity} for generic eccentric orbits.

\subsection{Equilibrium and dynamic tide}\label{sec:tide}

The tidal response in the static limit can be formally described by a transformation tensor
\begin{align}
Q_{ij} =\mathcal{T}_{i j \alpha \beta} \mathcal{E}^{\alpha \beta}\,.
\end{align}
For spherically symmetric object, this transformation tensor reduces to a single tidal Love numbers $\lambda$, such that
\begin{align}\label{eq:etide}
Q_{ij} = -\lambda \mathcal{E}_{ij}\,.
\end{align}

When the frequency for the tidal field $\omega$ is much less then the mode frequencies, we can approximate the solution of Eq.~\eqref{eq:qad} as 
\begin{align}
Q_n \approx -\mathcal{E}_Q /\omega^2_n\,,
\end{align}
which is often referred as the the equilibrium tide approximation. The total induced quadrupole moment is
\begin{align}
Q_{ij} =\sum_n Q^n_{ij} = -\frac{1}{2}\sum_n \frac{1}{\omega^2_n} \mathcal{O}^n_{ij} \mathcal{O}^{n*}_{ab} \mathcal{E}_{ab}\,.
\end{align}

Such expression should be compared with Eq.~\eqref{eq:etide}.
While it is not immediately clear why these two expressions are equivalent, they are guaranteed by the spherical symmetry of the star, and $\lambda$ can be obtained as
\begin{align}
\lambda &= \frac{1}{2}\sum_n \frac{1}{\omega^2_n ( \mathcal{E}_{ij}  \mathcal{E}^{ij})}  \mathcal{E}_{ij} \mathcal{O}^n_{ij} \mathcal{O}^{n*}_{ab} \mathcal{E}_{ab} \nonumber \\
& \approx \frac{4 \pi Q^2_\xi}{15 \omega^2_f}\,,
\end{align}
where in the second line we have kept the contribution from f-modes.
On the other hand,
for the case with
\begin{align}\label{eq:ep}
\mathcal{E} \sim \sum_{\alpha} b_\alpha e^{-i \Omega_\alpha t}
\end{align}
we have
\begin{align}\label{eq:qn}
Q^n \approx - \epsilon \sum_\alpha \frac{ b_\alpha e^{-i \Omega_\alpha t}}{\omega^2_n+(\gamma- i \Omega_\alpha)^2}\,.
\end{align}

The tidal response is frequency dependent, which is often referred as the dynamic tide. 
Certain higher harmonic frequency $\Omega_\alpha$ may be comparable or even larger than the f-mode frequency $\omega_f$.
Resonance occurs when $\Omega_\alpha \approx \omega_f$.

\section{Newtonian orbits}\label{sec:new}

In this section, we keep only the leading order Newtonian term in the Hamiltonian, and solve for the orbital evolution  in response to the tidal coupling. We 
also examine the tidal resonances in the low eccentricity limit, and discuss the criteria of detecting such resonances with current and future GW detectors.
For simplicity, we only include the mode evolution for one reference star, as it is straightforward to extend the analysis to oscillations of both stars in the BNS system.

The Newtonian Hamiltonian for the orbit is
\begin{align}
\mathcal{H}_0 = \frac{ p^2}{2 \mu} -\frac{M \mu}{r}= \frac{p^2_r}{2 \mu} +\frac{L^2_\phi}{ 2 m r^2} -\frac{M \mu}{r}\,,
\end{align}
with $\mu =M_1 M_*/M$ and $M = M_1 + M_*$. The unperturbed orbit can be characterized as 
\begin{align}\label{eq:tran}
r_{\rm orb} = & a (1-e \cos u) \,,\nonumber \\
l = & n (t-t_p )= u -e \sin u  \,,\nonumber \\
\phi-\phi_p = & 2 \arctan \left [ \sqrt{\left (\frac{1+ e}{1- e} \right )} \tan \frac{u}{2}\right ] \,,
\end{align}
with $n=\sqrt{M /a^3}$, $u$ being the mean anomaly, $l$ being the mean motion and $\phi$ being the true anomaly, and $t_p, \phi_p$ corresponding to the time and angle at the pericenter passage. We
further have energy and eccentricity given by
\begin{align}
E =-\frac{M \mu}{2 a}\,, \quad e^2 =1 +\frac{2 E J^2}{\mu^3 M^2 }\,.
\end{align}

Now according to Eq.~\eqref{eq:expand}, we expand the motion as
\begin{align}
r = r_{\rm orb} + \epsilon^2 \delta r,\quad L_\phi = L_{\rm orb} +\epsilon^2 \delta L_\phi\,.
\end{align}

The equations of motion (c.f. Eq.~\eqref{eq:eom}) become
\begin{align}\label{eq:eomrphi}
&\ddot \delta r +\frac{3 L^2_{\rm orb} \delta r}{\mu^2 r_{\rm orb}^4} -\frac{2 L_{\rm orb} \delta L_\phi}{\mu^2 r_{\rm orb}^3}-\frac{2 M \delta r}{r^3_{\rm orb}} = -\frac{\epsilon}{2 \mu} \frac{\partial \mathcal{E}^{ij}}{\partial r} Q_{ij}\,, \nonumber \\
& \frac{d \delta L_{\phi}}{d t} = -\frac{\epsilon}{2} \frac{\partial \mathcal{E}^{ij}}{ \partial \phi} Q_{ij} \,,
\end{align}
and the solution of the mode excitation is given by Eq.~\eqref{eq:dtide}.
For simplicity we neglect the initial oscillation by setting $Q_{\rm free} =0$.

\subsection{Dynamic tide}

The dynamical-tide excitation corresponds to the tidal response described by Eq.~\eqref{eq:dtide}.
For the ``background" trajectory described by Eq.~\eqref{eq:tran}, the tidal tensor generated by the companion NS with mass $M_*$, that acts on 
the reference NS, is \cite{Yang:2018bzx} (taking $\phi_p =0$ here)
 \begin{align}
 \mathcal{E}_{ij}= \frac{M_*}{r_{\rm orb}^3}\left [ \begin{array}{ccc}
 -\frac{1}{2}-\frac{3}{2}\cos 2\phi & \frac{3}{2}  \sin 2\phi & 0\\
 \frac{3}{2}  \sin 2\phi & \frac{-1}{2}+\frac{3}{2} \cos 2\phi & 0 \\
 0 & 0 & 1
   \end{array}
 \right ]
 \end{align}
 so that (restricting to $\ell=2$ subspace)
\begin{align}
\mathcal{E}^{m}_Q & = -\frac{ W_{2m} M_*}{r^3_{\rm orb}} Q_\xi e^{-i m \phi}\,.
\end{align}
Here $W_{lm}$ are defined in \cite{press1977formation}, with relevant components used in this work:
\begin{align}
W_{2,\pm 2} = \sqrt{\frac{3 \pi}{10}},\quad W_{2,0} = -\sqrt{\frac{\pi}{5}},\quad W_{2,\pm1} =0\,.
\end{align}

In order to solve the equations of motion in Eq.~\eqref{eq:eomrphi}, let us define \cite{rathore2005resonant}
\begin{align}
e^{i m \phi} (1+e \cos \phi)^n = \sum_{k \ge 0} c^m_{n,k} \cos k l+i s^m_{n,k} \sin k l\,.
\end{align}
Here $c^m_{n,k}, s^m_{n,k}$ are functions of the eccentricity $e$, which are proportional to the Hansen coefficients \cite{murray1999solar} (apart from a $(1-e^2)^n$ factor). They
can be obtained through

\begin{widetext}
\begin{align}\label{eq:c}
c^m_{n,k} =& \frac{1}{\pi (1+\delta_{k0})} \int^\pi_{-\pi } (1+e \cos\phi)^n \cos m \phi\cos k l \, dl \,\nonumber \\
= & \frac{(1-e^2)^n}{\pi (1+\delta_{k0})} \int^\pi_{-\pi } (1-e \cos u)^{1-n} \cos [k(u-e \sin u)] \cos \left \{ 2 m \arctan \left [ \sqrt{\left (\frac{1+ e}{1- e} \right )} \tan \frac{u}{2}\right ]\right \} d u\,\nonumber \\
= & c^{-m}_{n,k}\,,
\end{align}
and
\begin{align}\label{eq:s}
s^m_{n,k} =& \frac{1}{\pi (1+\delta_{k0})} \int^\pi_{-\pi } (1+e \cos\phi)^n \sin m \phi\sin k l \, dl \,\nonumber \\
= & \frac{(1-e^2)^n}{\pi (1+\delta_{k0})} \int^\pi_{-\pi } (1-e \cos u)^{1-n} \sin [k(u-e \sin u)] \sin \left \{ 2 m \arctan \left [ \sqrt{\left (\frac{1+ e}{1- e} \right )} \tan \frac{u}{2}\right ]\right \} d u\,\nonumber \\
= & -s^{-m}_{n,k}\,.
\end{align}
\end{widetext}

In terms of the Hansen coefficients, and according to Eq.~\eqref{eq:dtide}, the f-mode excitation becomes
($\gamma \rightarrow 0$)
 \newpage
 
\begin{align}\label{eq:qm}
& Q^m \approx  
-\frac{\epsilon} { \omega_f} \int^t_{-\infty} e^{-\gamma(t-t')} \sin \omega_f (t-t') \mathcal{E}_Q dt'\, \nonumber \\
&=-\frac{ W_{2m} M_* Q_\xi \epsilon }{  \omega_f }  \int^t_{-\infty} \frac{dt'}{r^3_{\rm orb}(t')} \sin \omega_f(t-t') e^{-i m \phi(t')} \nonumber \\
& \approx \frac{\epsilon  W_{2m} M_* Q_\xi }{   q^3}  \left ( \sum_k \frac{ c^m_{3,k} \cos k l}{-(k n)^2 +\omega^2_f}  -i \sum_k \frac{ s^m_{3,k} \sin k l}{-(k n)^2 +\omega^2_f} \right ) \,,
\end{align}
where in the last line we have adopted the adiabatic approximation (c.f. Eq.~\eqref{eq:qad}), and $q :=a(1-e^2)$.
The star is driven at integer harmonics of the orbital frequency, so that it is possible to have resonance crossing	
during the inspiral stage. For circular orbits, only the term with $k=2$ survives. In addition, the inspiral usually terminates
at the ISCO (Innermost-Stable-Circular-Orbit) frequency or the contact frequency of the two NSs, which are comparable 
or smaller than the f-mode frequency divided by two. Therefore it is difficult to observe a complete resonance during 
the inspiral stage for $k=2$. However, it has been shown \cite{Steinhoff:2016rfi} that the frequency dependence of the tidal love number is important
for describing motion in the late inspiral stage, which is essentially related to the $k=2$ resonance.

We can similarly decompose the driving terms in the equations of motion into a summation of harmonics:
\begin{widetext}
\begin{align}
&\mathcal{E}_{ij}Q_{ij}/2  = \sum_m \mathcal{E}^{m *}_Q Q_m \,, \nonumber \\
& = \sum_m \frac{\epsilon^2  W^2_{2m} M^2_* Q^2_\xi }{   q^6} \left ( \sum_{k'} c^m_{3, k'} \cos k' l  \sum_k \frac{ c^m_{3,k} \cos k l}{(k n)^2 -\omega^2_f} +\sum_{k'} s^m_{3, k'} \sin k' l  \sum_k \frac{ s^m_{3,k} \sin k l}{(k n)^2 -\omega^2_f} \right )\,.
\end{align}
and 
\begin{align}
& Q_{ij}/2 \frac{\partial \mathcal{E}_{ij}}{\partial r_{\rm orb}}  = \sum_m \frac{\partial \mathcal{E}^{m *}_Q}{\partial r_{\rm orb}} Q_m \,, \nonumber \\
& = -\sum_m \frac{\epsilon^2 3 W^2_{2m} M^2_* Q^2_\xi }{   q^7}\left ( \sum_{k'} c^m_{4, k'} \cos k' l  \sum_k \frac{ c^m_{3,k} \cos k l}{(k n)^2 -\omega^2_f} +\sum_{k'} s^m_{4, k'} \sin k' l  \sum_k \frac{ s^m_{3,k} \sin k l}{(k n)^2 -\omega^2_f}\right )\,,
\end{align}
and
\begin{align}\label{eq:qphi}
& Q_{ij}/2 \frac{\partial \mathcal{E}_{ij}}{\partial \phi}  = \sum_m \frac{\partial \mathcal{E}^{m *}_Q}{\partial \phi} Q_m \,, \nonumber \\
& =\sum_m \frac{\epsilon^2  m W^2_{2m} M^2_* Q^2_\xi }{   q^6} \left ( \sum_{k'} s^m_{3, k'} \sin k' l  \sum_k \frac{ c^m_{3,k} \cos k l}{(k n)^2 -\omega^2_f} -\sum_{k'} c^m_{3, k'} \cos k' l  \sum_k \frac{ s^m_{3,k} \sin k l}{(k n)^2 -\omega^2_f} \right ) \,.
\end{align}

As a result, we can determine Eq.~\eqref{eq:eomrphi} in an expansion of harmonics. We first write $\delta r$ and $\delta \phi$ as
\begin{align}
&\delta r =\sum_k b_k \cos k l , \quad \delta L_\phi = \sum_k g_k \cos k l \,.
\end{align}
There are no $\sin k l$ terms because of the absence of corresponding terms in the driving force.
The equations of motion become ($k \ge 0$)
\begin{align}\label{eq:meq}
&-k^2 n^2 b_k + \frac{3 L_{\rm orb}^2}{2 \mu^2 q^4} \sum_{h \ge 0} b_h (c^0_{4, |k-h|}+ c^0_{4, k+h}) -\frac{ M}{ q^3}  \sum_{m \ge 0} b_{h } (c^0_{3, |k-h|}+ c^0_{3, k+h}) -\frac{2 L_{\rm orb}}{\mu^2 q^3}  \sum_{h \ge 0} g_h (c^0_{3, |k-h|}+ c^0_{3, k+h}) \nonumber \\
& = \sum_m \frac{\epsilon^2 3 W^2_{2m} M^2_* Q^2_\xi }{  2 \mu q^7}\sum_{h \ge 0} \left [\frac{ c^m_{3,h} }{(h n)^2 -\omega^2_f} (c^m_{4, |k-h|} +c^m_{4, k+h}) + \frac{ s^m_{3,h} }{(h n)^2 -\omega^2_f} ({\rm Sign}(h-k)s^m_{4, |k-h|} +s^m_{4, k+h}) \right ] \,,\nonumber \\
& -k n g_k = - \sum_m \frac{\epsilon^2  m W^2_{2m} M^2_*  Q^2_\xi }{  2 q^6}  \sum_{h \ge 0} \left [\frac{ c^m_{3,h} }{(h n)^2 -\omega^2_f} ({\rm Sign}(k-h) s^m_{3, |k-h|} +s^m_{3, k+h}) + \frac{ s^m_{3,h} }{(h n)^2 -\omega^2_f} (-c^m_{3, |k-h|} +c^m_{3, k+h}) \right ]\,.
\end{align}
\end{widetext}
These equations can be solved in the matrix form. The $k=0$ piece of angular momentum shift $g_0$ is zero, because there is no DC angular momentum exchange between the orbit and the stars.

\subsection{Small eccentricity limit}

In the small eccentricity limit $e \ll 1$, we can take the leading order expansion of the Hansen coefficients in 
terms of $e$. They can be found in the Appendix A. 
In particular, for the terms showing up in Eq.~\eqref{eq:meq}, 
we always have $m=\pm2, 0$ and $n=3, 4$. The nonzero components that are proportional to $e$ are
\begin{align}
& c^0_{n,1} = n e, \quad s^0_{n,1}  =0,\quad \nonumber \\
& c^2_{3,3} = 3.5 e, \quad c^{-2}_{3,3} = 3.5 e, \nonumber \\
& s^2_{3,3} = 3.5 e, \quad s^{-2}_{3,3} =-3.5 e \nonumber \\
& c^2_{3,1} = -0.5 e, \quad c^{-2}_{3,1} = -0.5 e, \nonumber \\
& s^2_{3,1} = -0.5 e, \quad s^{-2}_{3,1} =0.5 e\,.
\end{align}
 The principle part that survives in the circular limit is
\begin{align}
& c^0_{n,0} =1,\quad s^0_{n,0} =0,\nonumber \\
& c^2_{n,2} =1,\quad s^2_{n,2} =1,\nonumber \\
& c^{-2}_{n,2} =1,\quad s^{-2}_{n,2} =-1\,.
\end{align} 

With these results we now try to solve Eq.~\eqref{eq:meq} in the small eccentricity limit.
The result is 
\begin{align}\label{eq:b0}
g_1 = & \epsilon^2 \frac{ W^2_{22} M^2_* Q^2_\xi}{ q^6} \frac{16 e n(n^2-4\omega^2_f)}{(9 n^2-\omega^2_f)(4 n^2- \omega^2_f)(n^2-\omega^2_f)}\,, \nonumber \\
b_0 = & \frac{\epsilon^2 3 M^2_* Q^2_\xi}{ a^4 \mu M} \left ( \frac{2 W^2_{22}}{4 n^2 -\omega_f^2} -\frac{W^2_{20}}{\omega^2_f}\right )\,, 
\end{align}
and we find that for $k=1$, the coefficient of $b_1$ becomes zero, which means that the $k=1$ term corresponds to the resonant frequency of the orbit. Any external driving frequency being the same as this frequency will formally make $b_1$ diverge. Physically what happens is that the coupling with star's internal degrees of freedom shifts the radial frequency, and make the orbit  precess. In order to fix this problem, we need to assign a different frequency $ n+\delta n$ to the radial motion with $\delta n \neq 0, \delta n/n \sim \mathcal{O}(\epsilon^2)$. The $k=1$ motion should also be absorbed into the background trajectory, with a redefinition of the eccentricity. By expanding the radial equations of motion in $\epsilon$ and keep terms linear in $\epsilon^2$, we see that
\begin{align}
& 6 e n^2 b_0+2n \mu \delta n e a
= \frac{2 L_{\rm orb}}{\mu a^3} g_1 \nonumber \\
&-\frac{\epsilon^2 3 M^2_* Q^2_\xi}{ a^7} \frac{4 \pi e (-36n^2 +100 n^4 \omega^2 -62 n^2 \omega^4+7 \omega^6)}{5 \omega^2 (\omega^2-9 n^2)(\omega^2-4n^2)}\,,
\end{align}
with $L_{\rm orb} \approx \mu n a^2$. As a result, we have
\begin{align}\label{eq:deltan}
\frac{\delta n}{n} &\approx -\frac{\epsilon^2 M^2_* Q^2_\xi}{\mu a^8 } \frac{6 \pi (18 n^6 -10 n^4 \omega^2 -12 n^2\omega^4+ \omega^6)}{5 \omega^2 n^2 (\omega^2-n^2)(\omega^2-4n^2)(\omega^2-9 n^2)}\,\nonumber \\
& \approx_{n \ll \omega} - \frac{6 \pi M^2_* Q^2_\xi}{5 \mu M \omega^2 a^5}\,.
\end{align}

All other terms ($b_k, g_k$) in the expansion series  are zero. With $\delta r$ and $\delta L_\phi$ known, it is then straightforward to compute
the tidal perturbation of the Hamiltonian. Both $\delta x$ and $\delta L_\phi$ are time-dependent, but the Hamiltonian perturbation is not.
Therefore we can pick any  time to evaluate $\delta \mathcal{H}$. It turns out that $l=\pi/2$ is a convenient choice.
\begin{widetext}
\begin{align}\label{eq:deltah}
\delta \mathcal{H}& =\epsilon^2 \frac{\partial \mathcal{H}_{\rm orb}}{\partial {\bf x}} \delta x + \epsilon^2 \frac{\partial \mathcal{H}_{\rm orb}}{\partial {\bf p}} \delta p +E_{\rm int}+E_{\rm mode} \nonumber \\
& = \frac{p_r \delta p_r}{a}+\frac{M \mu \delta r}{r^2_{\rm orb}} -\frac{L_{\rm orb}^2 \delta r}{\mu r^3_{\rm orb}}+\frac{\mathcal{E}_{ij} Q_{ij}}{2}+ \sum_{m} \frac{(\dot{Q}^m)^2+\omega^2_f (Q^m)^2}{2} \nonumber \\
& = \mu a^2 e^2\delta n n + \frac{M \mu b_0}{a^2} e^2 +\frac{\mathcal{E}_{ij} Q_{ij}}{2}+ \sum_{m} \frac{(\dot{Q}^m)^2+\omega^2_f (Q^m)^2}{2} \nonumber \\
& = \frac{\epsilon^2 M^2_* Q^2_\xi}{ a^6 } \left \{-\frac{2 \pi (4n^2-11n^2\omega^2+\omega^4)}{5 \omega^2 (\omega^2-4 n^2)^2} \right . \nonumber \\
& \left . +\frac{6 \pi e^2}{5} \frac{(648 n^{12}+2610 n^{10} \omega^2 -9470 n^8 \omega^4+8759 n^6\omega^6-2179 n^4 \omega^8+133 n^2 \omega^{10} +3 \omega^{12})}{\omega^2 (\omega^2-4n^2)^2 (\omega^4-10 n^2 \omega^2 +9n^4)^2} \right \}\,.
\end{align}

\end{widetext}
Similarly for the angular momentum:
\begin{align}
\delta J & = \epsilon^2 \frac{\partial \mathcal{J}_{\rm orb}}{\partial {\bf x}} \delta x + \epsilon^2 \frac{\partial \mathcal{J}_{\rm orb}}{\partial {\bf p}} \delta p  +\epsilon^2 J_{\rm mode} \nonumber \\
& = \epsilon^2 \delta L_\phi +\epsilon^2 J_{\rm mode}\,.
\end{align}
We notice that
\begin{align}
\frac{d J_{\rm mode}}{d t} =  \sum_m (\mathcal{E}_{2 j} \mathcal{O}^m_{1 j}- \mathcal{E}_{1 j} \mathcal{O}^m_{2 j}) Q_m =-\frac{d \delta L_\phi}{d t}\,,
\end{align}
so that  $\delta J$ is conserved. At $l=\pi/2$, $\delta L_\phi$ is zero (it has no $k=0$ component) and $J_{\rm mode}$ is given by \cite{Steinhoff:2016rfi}
\begin{align}\label{eq:jmode}
J_{\rm mode} & =\frac{\dot{Q}_{2 i} Q_{1 i} -\dot{Q}_{1i} Q_{2 i}}{\lambda \omega^2_f} \nonumber \\
& = \frac{1}{\lambda \omega^2_f} \sum_{m,m'} \dot{Q}_m Q_{m'} (\mathcal{O}^m_{2 l} \mathcal{O}^{m'}_{1 l} -\mathcal{O}^m_{1 l} \mathcal{O}^{m'}_{2 l}) \nonumber \\
& = \frac{1}{\lambda \omega^2_f} \frac{8 \pi i}{15} \frac{W^2_{22} M^2_\star Q^4_\xi}{q^6}(\dot{Q}_2 Q_{-2}-\dot{Q}_{-2} Q_2 ) \nonumber \\
& = \frac{16 \pi }{15} \frac{W^2_{22} M^2_\star n Q^4_\xi}{\lambda \omega^2_f q^6} \sum_k \frac{k c^2_{3,k} s^2_{3,k}}{((k n)^2 - \omega^2_f)^2}\nonumber \\
& = \frac{6\pi}{5} \frac{ M^2_\star n Q^2_\xi}{ q^6} \sum_k \frac{k c^2_{3,k} s^2_{3,k}}{((k n)^2 - \omega^2_f)^2}\, \nonumber \\
& \approx \frac{6\pi}{5} \frac{ M^2_\star n Q^2_\xi}{ a^6} \left \{ \frac{2}{(4n^2-\omega_f^2)^2} +e^2 \left [ \frac{1}{4(n^2-\omega_f^2)^2} \right . \right . \nonumber \\
&\left . \left .+\frac{12}{(4n^2-\omega_f^2)^2} +\frac{147}{(9n^2-\omega_f^2)^2} \right ]\right \}\,,
\end{align} 
where we have used the identification $\lambda =4 \pi Q^2_\xi/(15 \omega^2_f)$ for f mode.
In addition, because
\begin{align}
\frac{d \phi}{d t} = \frac{\partial \mathcal{H}_{\rm orb}}{\partial p_\phi}\,,
\end{align}
we can evaluate the perturbation of the angular frequency $n+\delta n_\phi$ due to the tidal interaction, such that
\begin{align}\label{eq:deltanphi}
\delta n_\phi &=\delta \left \langle \frac{\partial \mathcal{H}_{\rm orb}}{\partial p_\phi} \right \rangle \nonumber \\
& \approx \left \langle \frac{\delta L_\phi}{\mu r^2} -\frac{2 L_\phi \delta r}{\mu r^3} \right \rangle \nonumber \\
& = \frac{1}{\mu a^2} \sum_{k} g_{k} c^0_{2,k} -\frac{2L_\phi}{\mu a^3} \sum_{k} b_{k} c^0_{3,k}\,.
\end{align}

Because of the modification of the radial and angular frequencies $\delta n, \delta n_\phi$, and the shift of the trajectory ($\delta r, \delta L_\phi$),  the principle parts of the energy and angular momentum radiation are
correspondingly changed.

\begin{align}\label{eq:dpej}
\delta P^E_{\rm orb} & =-\frac{2}{5} \langle \delta \dddot{\mathcal{I}}_{jk} \dddot{\mathcal{I}}_{jk} \rangle\,,\nonumber \\
\delta P^J_{\rm orb} & = -\frac{2}{5} \epsilon_{3jk} \langle \delta \ddot{\mathcal{I}}_{jm} \dddot{\mathcal{I}}_{km} \rangle -\frac{2}{5} \epsilon_{3jk} \langle \ddot{\mathcal{I}}_{jm} \delta\dddot{\mathcal{I}}_{km} \rangle\,.
\end{align}

The explicit evaluation of these quantities are discussed in detail in Appendix \ref{app:3}.

\subsubsection{Sample evolution}

\begin{figure}
\includegraphics[width=8.4cm]{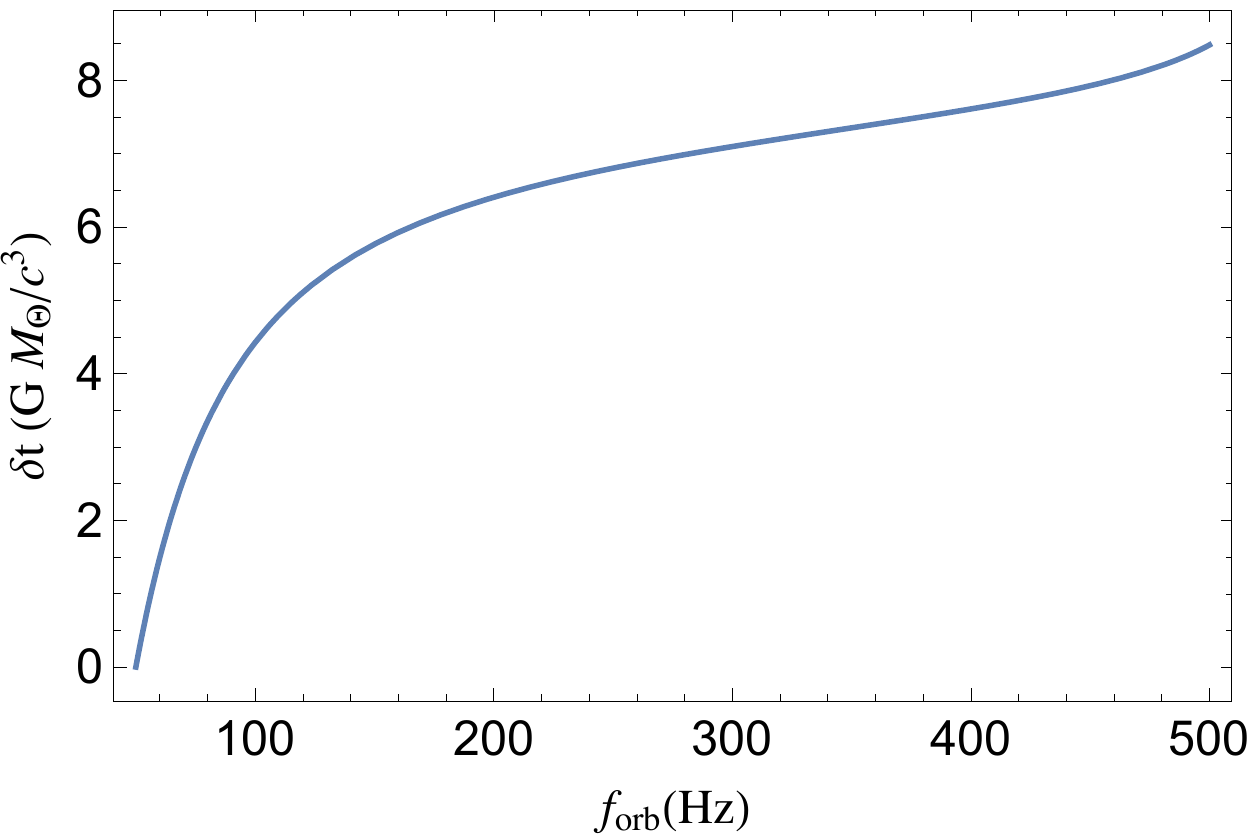}
\caption{$\delta t -f$ plot comparing two individual evolutions. Here $f_{\rm orb}$ is the orbital frequency, and $\delta t:= t_{\rm full} - t_{\rm part}$ is the difference in time between two evolutions up to a given frequency. The ``full" evolution incorporates the flux, energy and angular momentum formulas derived in Appendix \ref{ap:secepi} and Sec. \ref{sec:new}. The ``partial" evolution uses only previous known formulas, which do not contain correction at the $e^2 \lambda$ order. The initial eccentricity at $f_{\rm orb}=50 {\rm Hz}$ is $0.2$, the NS mass is $1.3 M_\odot$, and the compactness is assumed to be $0.16$.}
\label{fig:toplot}
\end{figure}

In order to  illustrate the effect of eccentric tidal terms in the flux, energy and angular momentum derived in Sec. \ref{sec:new} and Appendix. \ref{ap:secepi}, we 
take an equal-mass, $1.3 M_\odot+1.3 M_\odot$, binary NS system as an example. The star compactness is assumed to be $M_*/R_* =0.16$, and the initial eccentricity 
at $f_{\rm orb} = 50 {\rm Hz}$ is set to be $0.2$. We use this system to compare two separate evolution schemes. In the first evolution, we adopt  Eq.~\ref{eq:deo}, Eq.~\ref{eq:djorb}, Eq.~\ref{eq:pejorbe}, Eq.~\ref{eq:pecrosse},  Eq.~\ref{eq:pjcrosse}
and Eq.~\ref{eq:jmode} for a time-domain evolution to obtain a relation between $f_{\rm orb}$ and $t_{\rm full}$. In the second evolution, we drop the terms proportional to $e^2 \lambda$
in these formulas, and evolve the system again to obtain the function $t_{\rm part}(f_{\rm orb})$. The time difference up to certain orbital frequency, $\delta t$, is defined to be $t_{\rm full} -t_{\rm part}$.

The phase difference accumulated for the dominate $\ell=2,m=2$ mode, can be estimated as
\begin{align}
\delta \Psi_{22} & = 4 \pi \int^{f_{\rm up}}_{f_{\rm down}} f_{\rm orb} d \delta t
 \sim 0.06
\end{align}
for this particular evolution if we choose $f_{\rm down} = 50 {\rm Hz}, f_{\rm up} = 500 {\rm Hz}$.
Such phase difference is probably only observable for loud events in the third-generation detector era.

\subsection{Tidal resonance}

According to Eq.~\eqref{eq:qm}, when $k n \approx \omega_f$ we expect the f mode to be resonantly excited. As the orbit does not stay at resonance for infinite time because of radiative dissipation,  we need to account for the total phase shift across the resonance. In this case, we need to consider Eq.~\eqref{eq:dtide} taking into account the evolution of conserved quantities.

There are two major effects of these tidal resonances, which are intimately related to each other. At first, f mode is resonantly excited across the resonance,
and this free oscillation of f modes persists in the post-resonance stage.
This f-mode free oscillation also generates GW radiation at frequency $\omega_f/(2\pi)$.
Secondly, as the orbital energy and angular momentum transfers to the f-mode to support the free oscillation, the orbital motion after the resonance gradually deviates 
away from the one without the resonance, due to the change of the conserved quantities. In this case, the resulting orbital phase shift can also be determined.

\subsubsection{Free oscillation}

The  orbit evolution across resonance requires that there is  a free oscillation piece of $Q_m$, excited by the driving force.
In other words, the last line of Eq.~\eqref{eq:qm} needs to be modified in the near-resonance regime, which introduces a free oscillation piece.

If $kn =\omega_f + k \dot{n} (t-t_f)+\mathcal{O}(t-t_f)^2$, then the free oscillation piece of $Q_m $ is
\begin{widetext}
\begin{align}
Q^{\rm free}_m & = \frac{\epsilon  W_{2m} M_* Q_\xi }{ 2  q^3}  \left \{ \sum_k \frac{c^m_{3,k}}{\sqrt{k \dot{n}}} \left [ F_{\rm c}(\sqrt{k \dot{n}}(t-t_f)) \sin \omega_f t + F_{\rm s}(\sqrt{k \dot{n}}(t-t_f)) \cos \omega_f t \right ]\right \} \nonumber \\
&  -i \frac{\epsilon  W_{2m} M_* Q_\xi }{ 2  q^3}  \left \{ \sum_k \frac{s^m_{3,k}}{\sqrt{k \dot{n}}} \left [ F_{\rm s}(\sqrt{k \dot{n}}(t-t_f)) \sin \omega_f t + F_{\rm c}(\sqrt{k \dot{n}}(t-t_f)) \cos \omega_f t \right ]\right \}\,,
\end{align}
\end{widetext}
with the Fresnel functions 
\begin{align}
F_{\rm c}(x) =\int^x_{-\infty} \cos t^2 dt,\quad F_{\rm s} = \int^x_{-\infty} \sin t^2 dt\,.
\end{align}
We note that $F_{\rm c}(\infty) =F_{\rm s}(\infty) =\sqrt{\pi/2}$. 
The idea is that $Q^{\rm ad}_m$ smoothly transit to $Q^{\rm free}_m$ when $kn \rightarrow \omega_f$:

\begin{align}\label{eq:resq}
Q_m & = Q^{\rm ad}_m, \quad t<t_1 \nonumber \\
& = Q^{\rm free}_m, \quad t_1 <t <t_2 \nonumber \\
& = Q^{\rm ad}_m + \frac{\epsilon  \sqrt{\pi} W_{2m} M_* Q_\xi }{ 2  q^3} \nonumber \\
& \times  \left \{ \sum_k \frac{c^m_{3,k}-i s^m_{3,k}}{\sqrt{k \dot{n}}} \left [  \sin (\omega_f t +\pi/4)\right ]\right \},\quad t>t_2
\end{align}

It can be shown that the transition is smooth (results do not depend sensitively on $t_1, t_2$), for example, using the asymptotic behaviour discussed in Steinhoff $\&$ Hinderer. However, they miss this free oscillation piece in the post-resonance regime ($t>t_2$).

Free oscillation of f-modes has been observed in highly eccentric binaries \cite{gold2012eccentric,Yang:2018bzx} \footnote{For the implusive approximation in \cite{Yang:2018bzx} to hold, the eccentricity needs to be generally larger than $\sim 0.6$. As the eccentricity generally decays with time, the impulsive approximation may break down in the later part of the evolution, when $e<0.6$. }. While the orbital timescale may be long, the collective work of many harmonics gives rise to the impulsive interaction near the pericenter. In general, we need to consider $(\omega_f/n)$th order harmonic to reproduce this ``free oscillation" feather. If the orbit is highly eccentric, this task is computationally expensive, and it is preferable to apply the impulse approximation as in \cite{Yang:2018bzx}.

\subsubsection{Orbital phase shift}

According to Eq.~\eqref{eq:resq},  the amplitude  of the f-mode free oscillation gained after the $k n \approx \omega_f$ resonance is
\begin{align}
|Q_{m}| \sim \epsilon \frac{ W_{2m} M_* Q_\xi}{\omega_f q^3} \frac{\sqrt{\pi/k \dot{ n}}}{2}  \sqrt{(c^m_{3,k})^2+(s^m_{3,k})^2} \,.
\end{align}

The orbital energy decreases an extra amount across the resonance, due to the excitation of the free oscillation. As a result, the binary should merge faster than those without resonance crossing. 
In the small eccentricity limit, the only relevant tidal resonances are the associated with $k=2$ or $k=3$. The $k=2$ resonance, although being largest in amplitude, hardly takes place  in 
the inspiral part because the binary merger frequency is generally smaller than half of the f-mode frequency.  On the other hand, the mode amplitude for the $k=3$ resonance is proportional to the eccentricity.
The orbital frequency associated with this resonance, being one third of the f-mode frequency, could be smaller than the merger frequency.

Therefore, for $k=3$ resonance, the corresponding $(2,2)$ component phase shift of the gravitational waveform is \cite{Lai:1993di} (for equal mass binaries):
\begin{align}\label{eq:reswangle}
\Delta \phi & \sim -2 \times 4\pi \frac{t_D}{t_{\rm orb}} \frac{\omega^2_f \sum_m |Q_m|^2}{E_{\rm orb}} \nonumber \\
& \sim 0.68 \left ( \frac{f_{\rm mode}}{1.5 {\rm kHz}}\right )^{-2} \left ( \frac{\mathcal{Q}}{0.3}\right )^2 M^{-4}_{*1.3} R^2_{12} \left ( \frac{e}{0.02} \right )^2\,,
\end{align}
where the first factor of 2 comes from the mode energy for both NSs. Here $t_{\rm orb}$ is the orbital period at the resonance, $e$ is the eccentricity at the resonance, $t_D =f/\dot{f}$ at the resonance and $E_{\rm orb} = M \mu/(2 a)$ is the orbital energy at 
the resonance. The quantity $\mathcal{Q}$ is the dimensionless tidal overlap coefficient defined as 
\begin{align}
 \mathcal{Q} =\frac{Q_\xi}{ (M_1 R^2_*)^{1/2}}\,.
\end{align}

Notice that the orbital eccentricity $e$ decays with increasing orbital frequency $f$ due to GW radiation, with the scaling being approximately $e \approx f^{-19/18}$.
This means that a binary with eccentricity $\sim 0.02$ at $500$ Hz, would have eccentricity $\sim 0.2$ at $50$ Hz and $\sim 0.6$ at $20$ Hz. Such binaries may originate
from dynamical captures in globular clusters \cite{o2009gravitational} and multi-body dynamic evolutions. Despite of the recent developments, the rate of these channels are still subject to significant theoretical uncertainties.

The statistical phase error of an event with signal-to-noise ratio (SNR) $\rho$ is approximately $\sqrt{D-1}/\rho$ \cite{chatziioannou2017constructing}, with $D$ being the number of degrees of freedom in the parameter estimation.
To resolve a phase $\mathcal{O}(0.5)$ error with $D \sim \mathcal{O}(10)$, events with $\rho \ge 7$ is necessary. Note that this resonance happens in the late-inspiral stage, which is around $1$kHz for the quadrupole GW radiation, so that $\rho$ should represent the SNR of the waveform segment starting from the resonance and ending at merger \footnote{If the post-merger waveform can be accurately modelled, $\rho$ should also include the merger and post-merger SNR.}. For a $1.3 M_\odot+1.3 M_\odot$ binary NS system at $40 {\rm Mpc}$, the SNR for the inspiral waveform from $1$kHz to the plunge frequency ($\sim 1.4 {\rm kHz}$) is around 1.5 for Advanced LIGO design sensitivity \cite{martynov2016sensitivity} and around $20$ for Cosmic Explorer \cite{abbott2017exploring}. Therefore, such requirements are more likely satisfied with 
the third generation GW detectors, such as the Einstein Telescope, Cosmic Explorer, LIGO-HF  which also has decent mid-frequency sensitivity \cite{Miao:2017qot,martynov2019exploring}, etc.

If one (or both) NS(s) is a milli-second pulsar, the rotation frequency $f_s$ may be a couple of hundred Hz. As a result, the degeneracy between different 
f modes with different azimuthal number $m$ is broken. In particular, the frequency split is roughly  $0.5 m f_s$ \cite{lee1995nonradial}. If the NS spin counter-rotates with the binary, the mode with frequency $f_{\rm mode} -f_s$ is mostly excited; if   the NS spin co-rotates with the binary, the mode with frequency $f_{\rm mode} + f_s$ is mostly excited. Therefore for counter-rotating binaries, if the 
$f_{\rm mode} - f_s$ is smaller than the merger frequency, the $k=2$ tidal resonance is present in the inspiral stage \cite{Steinhoff:2016rfi}\footnote{Private communication with Yanbei Chen.}.
On the other hand, for the $k=3$ tidal resonance, the mode frequency  in Eq.~\eqref{eq:reswangle}  is $f_{\rm mode} -f_s$ and the  corresponding orbital frequency will be modified to $(f_{\rm mode}-f_s)/3$.

\section{Tidally Modified QK orbits}\label{sec4}

In the Post-Newtonian  limit, the motion of two gravitationally bounded points masses can be described by the Quasi-Keplerian orbit,
as a generalization of the Keplerian orbit in the Newtonian theory:
\begin{align}\label{eq:qk2}
& r=a_r (1-e_r \cos u)\,,\nonumber \\
& l= n'(t-t_p)= u- e_t \sin u +\sum l_{\rm PN}\,, \nonumber \\
& \frac{\phi-\phi_p}{K'} = v +\sum \phi_{\rm PN},
\end{align}
with
\begin{align}
\cos u =\frac{e_\phi+\cos v}{1 + e_\phi \cos v}.
\end{align}

Here the Post-Newtonian correction $l_{\rm PN}, \phi_{\rm PN}$ are functions 
of $u$ or $v$. Their detailed expressions, together with the expressions for
$e_r, e_\phi, e_t, n', K'$ can be found in \cite{blanchet2014gravitational}.
Unlike a Keplerian orbit, the radial and azimuthal frequencies of a QK trajectory are no longer degenerate.

Notice that if we are only interested in obtaining leading-order tidal effect in the Post-Newtonian expansion, we only need to  plug in the Newtonian part of $\mathcal{H}_{\rm orb}$ in Eq.~\eqref{eq:eom}, but sticking to Eq.~\eqref{eq:qk2}  for the prescription of the ``background" trajectory. In fact, we can further neglect $\phi_{\rm PN}$ and $l_{\rm PN}$ for this purpose because they contribute to oscillatory Post-Newtonian effect, as comparing to the secular Post-Newtonian effect encoded in $e_r, e_t, e_\phi, K', n'$.  Let us write $\delta r, \delta L_\phi$ as
\begin{align}
\delta r = & \sum_{m, k} b_{m,k} \cos (m K_\phi l + k l),\nonumber \\
\delta L_\phi = & \sum_{m, k} g_{m, k} \cos (m K_\phi l + k l)\,.
\end{align}
$K_\phi = K'+\delta K'$ is different from $K'$ because of the tidal correction.
Similarly $n'$ needs to be promoted to $n_t =n'+\delta n'$ by the tidal correction as well.
In the Newtonian, low-eccentricity limit, such mappings are described by
Eq.~\ref{eq:deltan} and Eq.~\ref{eq:deltanphi}.
Here $n'$ can also be determined by the equation of motion for the 
$k'=0, k=1$ component of orbital perturbations.

\subsection{Equations of motion}

The mode excitation of the star can still be determined from (in the non-resonant regime)
\begin{align}\label{eq:qm2}
& Q_m = 
-\epsilon /( \omega_f) \int^t e^{-\gamma(t-t')} \sin \omega_f (t-t') \mathcal{E}_Q dt'\, \nonumber \\
&=-\frac{ W_{2m} M_* Q_\xi \epsilon }{  \omega_f }  \int^t_{-\infty} \frac{dt'}{r^3_{\rm orb}(t')} \sin \omega_f(t-t') e^{-i m \phi(t')} \nonumber \\
& \approx\frac{\epsilon  W_{2m} M_* Q_\xi }{   a_r^3}  \left \{ \sum_{k,k'} \frac{ c^m_{3,k,k'} \cos (k+K' k') l}{-[(k+K' k' )n']^2  +\omega^2_f} \right . \nonumber \\
& \left . -i \sum_{k,k'} \frac{ s^m_{3,k,k'} \sin (k+K' k') l}{-[(k + K' k')n']^2 +\omega^2_f} \right \} \,,
\end{align}
where the QK-Hansen coefficients $c^m_{n,k,k'}, s^m_{n,k,k'}$, which depend on $e_{t, r, \phi}, K'$, can be defined by

\begin{align}\label{eq:qkhanc}
& c^m_{n,k,k'} =  c^{-m}_{n,k,k'} \nonumber \\
= & \lim_{T \rightarrow \infty}\frac{1}{T (1+\delta_{k0} \delta_{k'0})} \int^T_{-T } (1-e_r \cos u)^{-n} (1-e_t \cos u)\nonumber \\
& \times \cos [(k+K' k')(u-e_t \sin u)] \cos (m K' v) d u\,,
\end{align}
and
\begin{align}\label{eq:qkhans}
& s^m_{n,k,k'} =s^{-m}_{n,k,k'} \nonumber \\
= &  \lim_{T \rightarrow \infty} \frac{1}{T (1+\delta_{k0} \delta_{k'0})} \int^T_{-T } (1-e_r \cos u)^{-n} (1-e_t \cos u) \nonumber \\
& \times \sin [(k+K' k')(u-e_t \sin u)] \sin (m K' v) d u\,.
\end{align}

It it straightforward to check that the QK-Hansen coefficients $c^m_{n,k,k'}, s^m_{n,k,k'}$ reduce to $c^m_{n,k} (1-e^2)^{-n}, s^m_{n,k} (1-e^2)^{-n}$ if we take the limit $e_{r,t,\phi} = e, K'=1$.

Strictly speaking, Eq.~\ref{eq:eomrphi} is no longer valid as $\mathcal{H}_{\rm orb}$ in Eq.~\ref{eq:h} and Eq.~\ref{eq:eom} now contain higher-order Post-Newtonian corrections.
However, the leading-Post-Newtonian tidal perturbation in $\delta r, \delta L_\phi$ can be still obtained using Eq.~\ref{eq:eomrphi} with $Q_m$ given in Eq.~\ref{eq:qm2}. In
particular, the equations of motion for $\delta L_\phi$ imply that ($k_0 \equiv k_1 K'+k_2$)
\begin{align}
& \sum_{k_1, k_2} -k_0 n' g_{k', k} \sin (k_0 l)  = -\epsilon \sum_m \frac{\partial \mathcal{E}^{m *}_Q}{\partial \phi} Q_m  \nonumber \\
 = & \sum_m \frac{\epsilon^2  m W^2_{2m} M^2_* Q^2_\xi }{   a_r^6} \left ( \sum_{k_1 ,k_2} s^m_{3, k_1, k_2} \sin k_0 l  \sum_{k'_1,k'_2} \frac{ c^m_{3,k,k'} \cos k'_0 l}{-(k'_0 n)^2 +\omega^2_f} \right . \nonumber \\
-&\left . \sum_{k_1, k_2} c^m_{3, k_1, k_2} \cos k_0 l  \sum_{k'_1, k'_2} \frac{ s^m_{3,k'_1,k'_2} \sin k'_0 l}{-(k'_0 n)^2 +\omega^2_f} \right ) \,.
\end{align}

This implies that
\begin{align}
&-k_0 n' g_{k_0} =  \sum_m \frac{\epsilon^2  m W^2_{2m} M^2_*  Q^2_\xi }{  2 a^6}  \nonumber \\
&\times \sum_{k'_0} \left [\frac{ c^m_{3,k'_0} }{-(k'_0 n')^2 +\omega^2_f} ({\rm Sign}(k_0-k'_0) s^m_{3, |k_0-k'_0|} +s^m_{3, k_0+k'_0}) \right . \nonumber \\
&+ \left . \frac{ s^m_{3,k'_0} }{-(k'_0 n')^2 +\omega^2_f} (-c^m_{3, |k_0-k'_0|} +c^m_{3, k_0+k'_0}) \right ]\,.
\end{align}

Similarly for $\delta r$ we have
\begin{align}\label{eq:meq2}
&-k_0^2 n'^2 b_{k_0} + \frac{3 L_{\rm orb}^2}{2 \mu^2 a_r^4} \sum_{k'_0 \ge 0} b_{k'_0} (c^0_{4, |k_0-k'_0|}+ c^0_{4, k_0+k'_0}) \nonumber \\
&-\frac{ M}{ a_r^3}  \sum_{k'_0 \ge 0} b_{k'_0 } (c^0_{3, |k_0-k'_0|}+ c^0_{3, k_0+k'_0}) \nonumber \\
& = \frac{2 L_{\rm orb}}{\mu^2 a_r^3}  \sum_{k'_0 \ge 0} g_{k'_0} (c^0_{3, |k_0-k'_0|}+ c^0_{3, k_0+k'_0})  + \sum_{m} \frac{\epsilon^2 3 W^2_{2m} M^2_* Q^2_\xi }{  2 a_r^7} \nonumber \\
&\times \sum_{k'_0} \left [\frac{ c^m_{3,k'_0} }{(k'_0 n)^2 -\omega^2_f} (c^{m}_{4, |k_0-k'_0|} +c^m_{4, k_0+k'_0}) \right . \nonumber \\
&\left .+ \frac{ s^m_{3,k'_0} }{(k'_0 n)^2 -\omega^2_f} ({\rm Sign}(k'_0-k_0)s^m_{4, |k_0-k'_0|} +s^m_{4, k_0+k'_0}) \right ] \,.
\end{align}

The $k_0$ is selected within a 2-D lattice $(k_1,k_2)$, but with the requirement that $k_0 \ge 0$.
In the small eccentricity limit, the relevant components of QK-Hansen coefficients are (also see Appendix \ref{sec:han})
\begin{align}
& c^0_{n,1,0} = n e_r, \quad s^0_{n,1,0}  =0,\quad \nonumber \\
& c^2_{n,1,2} = \frac{n}{2} e_r +K'(e_t+e_\phi), \quad s^2_{n,1,2} = \frac{n}{2} e_r +K'(e_t+e_\phi) \,,\nonumber \\
& c^{-2}_{n,1,2} = \frac{n}{2} e_r +K'(e_t+e_\phi), \quad s^{-2}_{n,1,2} = -\frac{n}{2} e_r -K'(e_t+e_\phi) \,,
\end{align}
 and 
\begin{align}
& c^0_{n,0,0} =1,\quad s^0_{n,0,0} =0,\nonumber \\
& c^2_{n,0,2} =1,\quad s^2_{n,0,2} =1,\nonumber \\
& c^{-2}_{n,0,2} =1,\quad s^{-2}_{n,0,2} =-1\,.
\end{align}

Using these simplified coefficients, it is straight forward to work out the dominant components of $\delta r, \delta L_\phi$:

\begin{widetext}
\begin{align}\label{eq:pngb}
g_{1,0} = & \epsilon^2 \frac{ W^2_{22} M^2_* Q^2_\xi}{ q^6} \frac{8  {K'}^2 n[12 {K'}^2 n^2 (2 e_r-e_t-e_\phi)+n^2(e_t+e_\phi)-(6e_r+e_t+e_\phi)\omega^2_f]}{[(1+2 K')^2 n^2-\omega^2_f](4 {K'}^2 n^2- \omega^2_f)[(1-2 K')^2 n^2-\omega^2_f ]}\,, \nonumber \\
b_{0,0} = & \frac{\epsilon^2 3 M^2_* Q^2_\xi}{ a^4 \mu M} \left ( \frac{2 W^2_{22}}{4 {K'}^2 n^2 -\omega_f^2} -\frac{W^2_{20}}{\omega^2_f}\right ) \,.
\end{align} 
\end{widetext}

Similar to the Newtonian case,
the $k_1=0, k_2=1$ (so that $k_0 =1$) component of Eq.~\eqref{eq:meq2} determine the tidally induced orbital precession $\delta n' = n_t-n'$, as it is degenerate with the radial motion of the background trajectory. 
As emphasized earlier, Eq.~\eqref{eq:meq2} already neglects Post-Newtonian terms on its left hand side, which means $n'$ is in principle not consistent with this Newtonian equation without the right-hand-side forcing term.
However, as we are only interested in the leading Post-Newtonian order of $\delta n'$, keeping Newtonian order terms in the principle part should suffice.

\begin{widetext}
\begin{align}
&6 e_r {n'}^2 b_{0,0}+2 n' \delta n' e_r a_r \nonumber \\
= & \frac{2 L}{\mu^2 a^3}  \sum_{k'_0 \ge 0} g_{k'_0} (c^0_{3, |1-k'_0|}+ c^0_{3, 1+k'_0})
+   \sum_{m} \frac{\epsilon^2 3 W^2_{2m} M^2_* Q^2_\xi }{  2 a^7}\sum_{k'_0} \left [\frac{ c^m_{3,k'_0} }{(k'_0 n)^2 -\omega^2_f} (c^{m}_{4, |1-k'_0|} +c^m_{4, 1+k'_0}) \right . \nonumber \\
+ & \left . \frac{ s^m_{3,k'_0} }{(k'_0 n)^2 -\omega^2_f} ({\rm Sign}(k'_0-1)s^m_{4, |1-k'_0|} +s^m_{4, 1+k'_0}) \right ]\,.
\end{align}

The solution of the above equation, $\delta n'$, is given by

 \begin{align}
 \frac{\delta n'}{n' } & = - \frac{\epsilon^2 M^2_* Q^2_\xi}{e_r \mu {n'}^2 a_r^8  } \frac{3\pi}{20} \left \{ \frac{16 K' [12 {K'}^2 {n'}^2 (2 e_r-e_t-e_\phi)+(e_t+e_\phi){n'}^2 -(6 e_r +e_t+e_\phi)\omega^2_f]}{((1+2 K')^2 {n'}^2-\omega^2_f)(4 {K'}^2 {n'}^2- \omega^2_f)((1-2 K')^2 {n'}^2-\omega^2_f)} \right . \nonumber \\
&  \left . -\frac{48 {K'}^2 {n'}^2 (e_t +e_\phi)}{(1-4 {K'}^2)^2 {n'}^4-2 (1+4 {K'}^2) {n'}^2 \omega_f^2 +\omega_f^4} - 12 e_t \left [ -\frac{1}{\omega_f^2} +\frac{3}{4 {K'}^2 {n'}^2-\omega_f^2}\right ]\right . \nonumber \\
& \left . - e_r\left [ \frac{8}{\omega_f^2} +\frac{6}{{n'}^2- \omega_f^2} +\frac{9}{(1-2 K')^2 {n'}^2 -\omega_f^2} +\frac{24}{4 {K'}^2 {n'}^2 -\omega_f^2} +\frac{9}{(1+2K')^2 {n'}^2-\omega_f^2}\right ] \right \}\,,
 \end{align}
\end{widetext}
which reduces to Eq.~\eqref{eq:deltan} in the limit $e_{r,t,\phi} =e, K'=1$. On the other hand, the tidal perturbation of the angular frequency, $\delta n'_\phi$, can be evaluated
by
\begin{align}
\delta n'_\phi &=\delta (K' n') = \left \langle \frac{\partial \mathcal{H}_{\rm orb}}{\partial p_\phi} \right \rangle \nonumber \\
& \approx \left \langle \frac{\delta L_\phi}{\mu r^2} -\frac{2 L_\phi \delta r}{\mu r^3} \right \rangle \nonumber \\
& = \frac{1}{a^2_r} \sum_{k,k'} g_{k,k'} c^0_{2,k,k'} -\frac{2L_\phi}{a^3_r} \sum_{k,k'} b_{k,k'} c^0_{3,k,k'}\,.
\end{align}
According to Eq.~\eqref{eq:qkhanc} and  Eq.~\eqref{eq:qkhans}, only $k'=0$ pieces of $\delta r, \delta L_\phi$ would contribute to the right hand side of the above equation. This is because $m$ is zero in these integrations for the QK-Hansen coefficients, so that the product between harmonics with frequency being multiples of $1$ and frequency $k+K' k'$ should be zero, unless $k'=0$.
In the small eccentricity limit, it is just
\begin{align}
\delta n'_\phi = -\frac{2 L_\phi}{a^3_r} b_{0,0},
\end{align}
with $b_{0,0}$ giving in Eq.~\eqref{eq:pngb}.

\section{Conclusion}\label{sec5}

In this work, we have discussed the trajectory model of an eccentric binary neutron star system, that evolves under the the influence
of dynamic tidal interaction and gravitational radiation. This formalism is suitable for both Newtonian and Post-Newtonian description of
the conserved dynamics.  We focus more on the Newtonian description in the present study,
as the leading order tidal correction can already be obtained in the Newtonian framework.

Within eccentric orbits, the direction of tidal bulges on the stars generally do not point to the companion star.
This is different from circular binaries, where the tidal bulges always point to each other, even with the consideration
of dynamic tide. As a result, the stars oscillatorily exchange orbital angular moment and mode angular momentum within
orbital timescales. In addition, the energy and angular momentum fluxes are also modified by the beating between the orbital
quadrupole moment and the star quadrupole moment. For binaries with eccentricity $e \sim 0.2$ at $f_{\rm orb} = 50$ Hz,
the eccentric-tidal effect on the $22$ mode radiation is only detectable by third-generation gravitational-wave detectors.

Eccentric tidal interaction also leads to tidal resonances in the inspiral stage. For circular binary, this resonance happens at
$\omega_{\rm orb} = \omega_f/2$, which is likely higher than the merger frequency of the binary, depending on the star equation of state.
For eccentric orbits, the tidal resonances show up at $\omega_{\rm orb} = \omega_f/k$ with $k \ge 2$, although the high order resonances
are generally weaker for low-eccentricity binaries. We have analyzed the first eccentric tidal resonance, which shows up at $\omega_{\rm orb} = \omega_f/3$. We argue
that it can be observed in the third-generation detector era. As the GW detectors are continuously  improving in sensitivity, there is a growing interest to characterize
the gravitational-wave radiation at the late-inspiral stage and the merger/post-merger stage, as a way to probe the neutron star physics beyond the information about tidal love number  \cite{Miao:2017qot}.

In order to build a Post-Newtonian waveform model for binary neutron stars at arbitrary eccentricity, one needs to solve the matrix equation for the Fourier components of $\delta r, \delta L_\phi$,
which in turn affects the gravitational-wave radiation of the orbit. For a frequency-domain description, additional difficulty arises in the analytical transformation from time-domain waveform to frequency-domain waveform under the stationary phase approximation, which have been discussed for eccentric binary black holes \cite{moore2018fourier} in the Newtonian limit. An alternative route is to use the Effective-One-Body framework for the system, for which the time-domain waveform are solved up to $1.5$ Post-Newtonian order \cite{Hinderer:2017jcs}. 
We shall leave the construction and validation of the eccentric binary neutron star waveform to future work.

At last, although the discussion here is presented on binary neutron star systems, the formalism is still valid for eccentric black hole-neutron star binaries. As the mass ratios for these systems 
are expected to be larger than those of binary neutron stars, except for low-mass black hole and neutron star binaries formed in more exotic astrophysical channels \cite{Yang:2017gfb}, it is
reasonable to expect very different initial eccentricity distribution at LIGO band even for dynamically formed binaries. 

\acknowledgements The author thank William East, Vasileios Paschalidis, Frans Pretorius, John Ryan Westernacher-Schneider
for interesting discussions.   This research was supported by NSERC and in part by the Perimeter
Institute for Theoretical Physics. Research at Perimeter Institute is
supported by the Government of Canada through the Department of
Innovation, Science and Economic Development Canada and by the
Province of Ontario through the Ministry of Research, Innovation and
Science.

\appendix

\section{Hansen coefficients}\label{sec:han}

In general, $c^m_{n,k}$ and $s^m_{n,k}$ can be obtained through numerical integration, based on Eq.~\eqref{eq:c} and Eq.~\eqref{eq:s}. However, in special 
cases they satisfy the following relations:
\begin{align}
e^{i \phi}  = & -e + \sum^\infty_{p=1} \left [ \frac{2(1-e^2)}{e} J_p (p e) \cos p l \right . \nonumber \\
&\left . +i \sqrt{1-e^2}\frac{2}{p} \frac{d J_p(p e)}{d e} \sin p l\right ]\,,
\end{align}
and 
\begin{align}
\cos \phi (1+e \cos \phi)^2 & =(1-e^2)^2\sum^\infty_{p=1} p [J_{p-1}(p e) - J_{p+1}(p e)] \cos p l\,,\nonumber \\
\sin \phi (1+e \cos \phi)^2 & =(1-e^2)^2\sum^\infty_{p=1} p [J_{p-1}(p e) +J_{p+1}(p e)] \sin p l\,,
\end{align}
where $J_k(x)$ is the Bessel function of the first kind. Some other useful relations are
\begin{align}
\frac{1}{1- e \cos u} = 1 +\sum^\infty_{p=1} 2 J_p(p e) \cos p l\,,
\end{align}
and 
\begin{align}
\cos k u =& -\frac{e (1-\delta_{k 0})}{2} \nonumber \\
&+\sum^\infty_{p=1} \frac{k}{p} [J_{p-k}(pe)-J_{p+k}(p e)] \cos p l\,.
\end{align}

In the small eccentricity limit, $c^m_{n,k}, s^m_{n,k}$ become

\begin{widetext}
 \begin{align}
c^m_{n,k}= & \frac{\delta_{k,m}+\delta_{k,-m}}{1+\delta_{k, 0}}+ e(n-1)\frac{\delta_{k,m-1}+\delta_{k,1-m}+\delta_{k,-m-1}+\delta_{k,1+m}}{2(1+\delta_{k, 0})} +e k \frac{-\delta_{k,m-1}+\delta_{k,1-m}-\delta_{k,-m-1}+\delta_{k,1+m}}{2(1+\delta_{k, 0})} \nonumber \\
&-e m \frac{\delta_{k,m-1}+\delta_{k,1-m}-\delta_{k,-m-1}-\delta_{k,1+m}}{2(1+\delta_{k, 0})}\,,
 \end{align}
 and
 \begin{align}
s^m_{n,k}= & \frac{\delta_{k,m}-\delta_{k,-m}}{1+\delta_{k, 0}}+ e(n-1)\frac{\delta_{k,m-1}-\delta_{k,1-m}-\delta_{k,-m-1}+\delta_{k,1+m}}{2(1+\delta_{k, 0})} -e k \frac{\delta_{k,m-1}+\delta_{k,1-m}-\delta_{k,-m-1}-\delta_{k,1+m}}{2(1+\delta_{k, 0})} \nonumber \\
&+e m \frac{-\delta_{k,m-1}+\delta_{k,1-m}-\delta_{k,-m-1}+\delta_{k,1+m}}{2(1+\delta_{k, 0})}\,.
 \end{align}

Similarly, for the QK-Hansen coefficients we define in Eq.~\eqref{eq:qkhanc} and  Eq.~\eqref{eq:qkhans}, the low eccentricity limit is given by
(with $\kappa :=k + K' k'$)

 \begin{align}
 c^m_{n,k,k'} = &\frac{\delta_{k0}(\delta_{m k'}+\delta_{-m, k'})}{1+\delta_{k0}\delta_{k'0}} + \frac{m K' e_\phi (\delta_{k' m}-\delta_{k',-m})}{2(1+\delta_{k0}\delta_{k'0})}(\delta_{k 1}-\delta_{k,-1})
 +\frac{(n e_r-e_t)(\delta_{m k'}+\delta_{-m, k'})(\delta_{k 1}+\delta_{k,-1})}{2(1+\delta_{k0}\delta_{k'0})} \nonumber \\
 + & \frac{\kappa e_t (\delta_{m k'}+\delta_{-m, k'})(\delta_{k 1}-\delta_{k,-1})}{2(1+\delta_{k0}\delta_{k'0})}\,,
 \end{align}
 and
  \begin{align}
 s^m_{n,k,k'} = &\frac{\delta_{k0}(\delta_{m k'}-\delta_{-m, k'})}{1+\delta_{k0}\delta_{k'0}} + \frac{m K' e_\phi (\delta_{k' m}+\delta_{k',-m})}{2(1+\delta_{k0}\delta_{k'0})}(\delta_{k 1}-\delta_{k,-1})
 +\frac{(n e_r-e_t)(\delta_{m k'}-\delta_{-m, k'})(\delta_{k 1}+\delta_{k,-1})}{2(1+\delta_{k0}\delta_{k'0})} \nonumber \\
 + & \frac{\kappa e_t (\delta_{m k'}-\delta_{-m, k'})(\delta_{k 1}-\delta_{k,-1})}{2(1+\delta_{k0}\delta_{k'0})}\,.
 \end{align}
 \end{widetext}

\section{Precession in the large eccentricity case}

 Consider a 1-D problem with Hamiltonian $\frac{p^2}{2 \mu} + V(x) $, the period is
\begin{align}
T = & 2 \int^{x_{\rm max}}_{x_{\rm min}} \frac{\sqrt{2 \mu} dx}{\sqrt{E - V(x)}} \nonumber \\
& =4 \frac{\partial}{\partial E} \int^{x_{\rm max}}_{x_{\rm min}} \sqrt{2 \mu} \sqrt{E - V(x)} dx\,.
\end{align}

If $V(x)$ is perturbed to $V(x) + \delta V(x)$, $T$ is perturbed as
\begin{align}
\delta T & = - 2\frac{\partial}{\partial E} \int^{x_{\rm max}}_{x_{\rm min}} \frac{\sqrt{2 \mu} \delta V dx}{\sqrt{E - V(x)}} \nonumber \\
& = - 2\frac{\partial}{\partial E} \int^{x_{\rm max}}_{x_{\rm min}} \delta V dt\,.
\end{align}

Now if the relevant forcing terms in the equation of motion for $\delta r$ (the right hand side of Eq.~\eqref{eq:meq2}) can be expressed as
$\sum_k F_k \cos k l$,
we have
\begin{align}
\delta V & = -\int \sum_k F_k \cos k l dr \nonumber \\
& = - a_r e_r \sum_k F_k  \int^u_{\pi/2} \cos kl \sin u d u\,,
\end{align} 
 so that
 \begin{align}
 \delta T =&  \frac{2 a_r e_r}{n'} \frac{\partial}{\partial E}  \int^\pi_{\pi} du (1-e_t \cos u) \nonumber \\
  & \times \int^u_{\pi/2}  \sum_k F_k \cos kl \sin u' d u' \,,
 \end{align}
 which gives the tidal-induced frequency shift, as $\delta T/T = - \delta n'/n'$.

 \section{e-$\varpi$ representation}\label{ap:secepi}

 In the main text (c.f. Section \ref{sec:new}) the tidal perturbation is evaluated with respect to constant semi-major axis $a$, and the radial and azimuthal frequencies deviate from the 
 Keplerian frequency. In practise, it is more convenient to discuss the modification at fixed azimuthal frequency $n_\phi$, as this is more suitable for constructing the frequency-domain 
 waveforms. Therefore we shall rewrite some of the key results in Section \ref{sec:new} with respect to fixed $n_\phi$. For convenience we define $\varpi :=n_\phi$ and $x := \varpi/\omega_f$.

 In order to ensure constant $\varpi$, the radius has to be further shift by $\delta a$ with (c.f. Eq.~\eqref{eq:deltanphi})
 \begin{align}
 \frac{\delta a}{a} =& -\frac{4}{3} \frac{(1-e^2)^{1/2} b_0}{a} \nonumber \\
 & +\frac{4}{3} \frac{e g_1}{\mu a^2 \varpi}\,\nonumber \\
 & = \frac{\lambda M_*}{M_1 a^5} \left [12 \frac{1-x^2}{1-4 x^2}   +\frac{3e^2}{2} \frac{31+42 x^2+371x^4-144 x^6}{(1-9x^2)(1-4 x^2)(1-x^2)}\right ]\,,
 \end{align}
 where in the last line we have extended the expression in Eq.~\eqref{eq:b0} to include order $\mathcal{O}(e^2)$ corrections:
 \begin{align}
 b_0 = & \frac{9  M_* \lambda}{4 M_1 a^4 } (1-3 e^2)\left ( \frac{3}{4x^2-1} -1\right ) \nonumber \\
 & + \frac{9  M_* \lambda e^2}{8 M_1 a^4 } \frac{(396 x^6-919 x^4+282 x^2-59)}{(1-9x^2)(1-4x^2)(1-x^2)}\,.
 \end{align}
 
 At constant $\varpi$, the tidally induced radius shift is
 \begin{align}
 b'_0 = & b_0+\delta a =\frac{3 M_*}{4 M_1} \frac{\lambda}{a^4} \left ( \frac{3}{1-4x^2} +1\right ) \, \nonumber \\
 & + \frac{3 e^2 M_*}{8 M_1} \frac{\lambda}{a^4} \frac{(-36 x^6+95 x^4+222 x^2+19)}{(1-9x^2)(1-4x^2)(1-x^2)}\,,
 \end{align}
 and $b'_1$ is now $-e \delta a$. The radial frequency shift is
 \begin{align}\label{eq:nr}
 \delta n_r & =\delta n-n \frac{3 \delta a}{2 a} \nonumber \\
 &=  \frac{9  M_* \lambda n}{2 M_1 a^5 } \frac{(-5+56x^2-66x^4+18x^6)}{(1-9x^2)(1-4x^2)(1-x^2)} \,.
 \end{align}
 The energy shift at constant $\varpi$ is $\delta \mathcal{H}$ in Eq.~\eqref{eq:deltah} plus modification due to $\delta a, \delta n_r$, with $g_0/L_\phi =\delta a /(2a)$:
 \begin{widetext}
 \begin{align}\label{eq:deo}
 \delta E(\varpi) = & \mu a^2 e^2 n_r \delta n_r+\mu a \delta a e^2 n^2+ \frac{M e^2 \mu b'_0}{a^2}+\frac{L_\phi \delta L_\phi}{\mu a^2} +\frac{\mathcal{E}_{ij} Q_{ij}}{2}+ \sum_{m} \frac{(\dot{Q}^m)^2+\omega^2_f (Q^m)^2}{2} \nonumber \\
 = & \mu (M \varpi)^{2/3}\frac{ \lambda M_*}{ M_1 a^5}\left \{ \frac{9}{2}\frac{1-3 x^2+4 x^4}{(1-4 x^2)^2} +\frac{3e^2}{4} \frac{29+234 x^2 -6636 x^4+29878 x^6-34773 x^8+11700x^{10}+2592x^{12}}{(1-4x^2)^2(1-x^2)^2 (1-9x^2)^2}\right \}\,.
\end{align}
 \end{widetext}
 
 Similarly the orbital-averaged modification in orbital angular momentum $\delta J_{\rm orb}$ is just $g_0$, which is
 \begin{align}\label{eq:djorb}
 \delta J_{\rm orb} =&  \frac{\mu \varpi  \lambda M_*}{2 M_1 a^3} (1-e^2)^{1/2} \left [12 \frac{1-x^2}{1-4 x^2}  \right . \nonumber \\
&\left.  +\frac{3e^2}{2} \frac{31+42 x^2+371x^4-144 x^6}{(1-9x^2)(1-4 x^2)(1-x^2)}\right ]\,.
 \end{align}

 \section{Evaluation of the fluxes}\label{app:3}
 
 In the Newtonian limit, we first consider a ``background" trajectory, which can be written as
 \begin{align}\label{eq:tran}
r_{\rm orb} = & a (1-e \cos u) \,,\nonumber \\
l = & n_r (t-t_p )= u -e \sin u  \,,\nonumber \\
\phi-\phi_p = & \frac{n_\phi}{n_r} v \,.
\end{align}
 
 Adopting the convention of  \cite{poisson2014gravity}, 
 we define the vector ${\bf n}, {\bf \xi}$, which are the unit vectors $\hat{r}, \hat{\phi}$. We also define $\Pi_1 = {\bf n} {\bf n}, \,\Pi_2 ={\bf n} {\bf \xi}+{\bf \xi} {\bf n}, \,\Pi_3 = {\bf \xi} {\bf \xi}$. We have $I_{ij} = \mu x_i x_j$, $\mathcal{I}_{ij} = I_{ij}-I \delta_{ij}/3$ and
 \begin{align}
 \dot{\Pi}_1 = \dot{\phi} \Pi_2, \quad \dot{\Pi}_2 =2 \dot{\phi} (\Pi_3 -\Pi_1),\quad \dot{\Pi}_3 =-\dot{\phi} \Pi_2\,,
 \end{align}
 and
 \begin{align}
 \ddot{I_{ij} } = & (2 \dot{r}^2-2r^2 \dot{\phi}^2+2 r \ddot{r})\Pi_1\nonumber \\
 & + (4 r \dot{r} \dot{\phi} +r^2 \ddot{\phi}) \Pi_2 +2r^2 \dot{\phi}^2 \Pi_3 \,,
 \end{align}
 and
 \begin{align}
 \dddot{I_{ij} } = &[12 r \dot{r} (\dot{\phi})^2 +6\dot{r}\ddot{r}-6r^2\dot{\phi}\ddot{\phi}+2r\dddot{r}] \Pi_1\nonumber \\
 & +\{6 \dot{r}^2 \dot{\phi} +6 r \dot{r} \ddot{\phi} +r(6 \dot{\phi} \ddot{r}+r[-4\dot{\phi}^3+\dddot{\phi})] \} \Pi_2 \nonumber \\
 & +6(2 r \dot{r} \dot{\phi}^2+r^2 \dot{\phi} \ddot{\phi}) \Pi_3\,.
 \end{align}

 Therefore we have the associated 
 fluxes for this orbital trajectory being
 \begin{widetext}
 \begin{align}
 P^E_{\rm back} & =-\frac{1}{5} \langle  \dddot{\mathcal{I}}_{jk} \dddot{\mathcal{I}}_{jk} \rangle = -\frac{1}{5} \left  \langle  \dddot{I}_{jk} \dddot{I}_{jk} -\frac{1}{3} \dddot{I}_{jj} \dddot{I}_{kk}\right \rangle \nonumber \\
 & \approx-\frac{32}{5}M^{4/3} \mu^2 \varpi^{10/3} \left [\frac{1}{(1-e^2)^{7/2}}\left ( 1+\frac{73}{24} e^2+\frac{37}{96}e^4 \right ) + \right . \nonumber \\
 & \left . -\frac{279 e^2 M_* \lambda }{8 M_1 a^5 } \frac{(-5+56x^2-66x^4+18x^6)}{(1-9x^2)(1-4x^2)(1-x^2)}\right ]\,,
  \end{align}
 where we have identified $n_\phi$ with $\varpi$ and used the expansion of $n_r$ in  Eq.~\eqref{eq:nr}. Similarly, the expansion of angular momentum
 flux for the background trajectory is
 \begin{align}
 P^J_{\rm back}  = & -\frac{2}{5} \epsilon_{3jk} \langle \ddot{\mathcal{I}}_{jm} \dddot{\mathcal{I}}_{km} \rangle \nonumber \\
 &= -\frac{32}{5} M^{4/3} \mu^2 \varpi^{7/3} \left [ \frac{1}{(1-e^2)^2}\left ( 1+\frac{7}{8} e^2 \right )  \right. \nonumber \\
 & \left . +\frac{9 e^2 M_* \lambda }{8 M_1 a^5 } \frac{(70-784x^2+924 x^4-252 x^6)}{(1-9x^2)(1-4x^2)(1-x^2)} \right ]\,.
 \end{align}
\end{widetext}
 
The physical trajectory is deformed from the background trajectory by $r \rightarrow r+b'_0$ and $\phi \rightarrow \phi+\delta \phi$, with $\delta \dot{\phi} = g_1 \cos l/(\mu a^2 )$. 
 As a result, we can apply Eq.~\ref{eq:dpej} and obtain
 \begin{widetext}
 \begin{align}\label{eq:pejorbe}
 P^E_{\rm orb} & = -\frac{32}{5}M^{4/3} \mu^2 \varpi^{10/3} \left [\frac{1}{(1-e^2)^{7/2}}\left ( 1+\frac{73}{24} e^2+\frac{37}{96}e^4 \right ) +12 \frac{ M_* \lambda }{ M_1 a^5 }\frac{1-x^2}{1-4x^2}\right . \nonumber \\
 & \left . +\frac{ 3e^2 M_* \lambda }{8 M_1 a^5 } \frac{(2328-8626x^2+13339x^4-5049x^6)}{(1-9x^2)(1-4x^2)(1-x^2)}\right ]\,,\\
 P^J_{\rm orb} & = -\frac{32}{5} M^{4/3} \mu^2 \varpi^{7/3} \left [ \frac{1}{(1-e^2)^2}\left ( 1+\frac{7}{8} e^2 \right ) +12 \frac{ M_* \lambda }{ M_1 a^5 }\frac{1-x^2}{1-4x^2} \right. \nonumber \\
 & \left . +\frac{3 e^2 M_* \lambda }{4 M_1 a^5 } \frac{(941-1899x^2+3286 x^4-1260 x^6)}{(1-9x^2)(1-4x^2)(1-x^2)} \right ]\,.
  \end{align}
 \end{widetext}
 
 The total energy and angular momentum fluxes are approximately the simulation between orbital flux $P^{E,J}_{\rm orb}$ and $P^{E,J}_{\rm orb}$, as the modal fluxes 
 are higher order in $\epsilon$. In the small eccentricity limit, they are
 \begin{align}\label{eq:pecrosse}
 P^E_{\times} & =  -\frac{1}{5} M^{4/3} \mu^2 \varpi^{10/3}\times \frac{ M \lambda }{ M_1 a^5 } \nonumber \\
& \times \left [\frac{192}{1-4x^2} +e^2\left (\frac{11}{2(1-x^2)}+\frac{64}{1-4x^2} +\frac{15309}{2(1-9x^2)} \right ) \right ]\,,
 \end{align}
 and
  \begin{align}\label{eq:pjcrosse}
 P^J_{\times} & =  -\frac{1}{5} M^{4/3} \mu^2 \varpi^{7/3}\times \frac{ M \lambda }{ M_1 a^5 } \nonumber \\
& \times \left [\frac{192}{1-4x^2} +e^2\left (\frac{9}{1-x^2}+\frac{216}{1-4x^2} +\frac{5103}{1-9x^2} \right ) \right ]\,.
 \end{align}
 
 Note that Eq.~\eqref{eq:pecrosse}, Eq.~\eqref{eq:pjcrosse} together with Eq.~\eqref{eq:pejorbe} produce consistent result with
 Eq.~8 in \cite{Flanagan:2007ix} in the circular limit. In particular, when $e=0$, we have 
 \begin{align}
 P^E_{\rm orb}  = \varpi P^J_{\rm orb} ,\quad  P^E_{\times} = \varpi  P^J_{\times}\,.
 \end{align}
 Using Eq.~\eqref{eq:deo}, Eq.~\eqref{eq:djorb} and Eq.~\eqref{eq:jmode}, it is straightforward to check that
  \begin{align}
 \frac{d {E}(\varpi)}{d \varpi} = \varpi \left [\frac{d {J}_{\rm orb}(\varpi)}{d \varpi} + \frac{d {J}_{\rm mode}(\varpi)}{d \varpi}\right ]\,,
 \end{align}
 or
 \begin{align}
 \dot{E}(\varpi) = \varpi [\dot{J}_{\rm orb}(\varpi) + \dot{J}_{\rm mode}(\varpi)]\,,
 \end{align}
 so that circular binaries remain circular even with the inclusion of dynamic tide.
 
\bibliography{master}
\end{document}